# Enhanced reversible barocaloric effect at low pressure in neopentyl plastic crystal solid solutions

*Frederic Rendell-Bhatti[*1], Melony Dilshad[2], Celine Beck[1], Markus Appel[3], Alba Prats[4], Eamonn T. Connolly[5], Claire Wilson[6], Lewis Giannelli[1], Pol Lloveras[4], Xavier Moya[2], David Boldrin[1], Donald A. MacLaren[1]*

[1]SUPA, School of Physics & Astronomy, University of Glasgow, Glasgow G12 8QQ, United Kingdom

[2]Department of Materials Science & Metallurgy, University of Cambridge, Cambridge, CB3 0FS, United Kingdom

[3]Institut Laue Langevin, 71 Avenue des Martyrs, 38000 Grenoble, France

[4]Group of Characterization of Materials, Department of Physics and Barcelona Research Center in Multiscale Science and Engineering, Universitat Politecnica de Catalunya, Av. Eduard Maristany 16, 08019, Barcelona, Spain

[5]Diamond Light Source, Diamond House, Harwell Science and Innovation Campus, Didcot, Oxfordshire, OX11 0DE, United Kingdom

[6]School of Chemistry, University of Glasgow, Glasgow G12 8QQ, United Kingdom

**Corresponding author email: fred.rendell@glasgow.ac.uk*

## Abstract

The discovery of colossal barocaloric effects in neopentyl glycol (NPG) makes plastic crystals promising candidates for solid-state refrigerants with lower environmental impact than vapour compression fluids. Optimising operational temperatures and low-pressure operability remains challenging without compromising thermodynamic parameters. Here, we implement a strategy to improve the viability of NPG derivatives as barocaloric refrigerants. We blend pentaglycerine (PG) with NPG to lower the phase transition temperature, then dope the blend with 2% pentaerythritol (PE) to improve transition reversibility. In comparison with NPG under the same conditions, this ternary system has a seven-fold increase in reversible isothermal entropy change ($|\Delta S_{\mathrm{it,rev}}|$ = 13.4 J kg$^{-1}$ K$^{-1}$) and twenty-fold increase in operational temperature span ($\Delta T_{\mathrm{span}}$ = 18 K) at pressures of 1 kbar. Synchrotron x-ray diffraction and quasielastic neutron scattering reveal structural and dynamical effects that broaden the temperature range of the first-order phase transition due to intermolecular hydrogen bond network disruption by the molecular dopants. We propose that exploiting the compositional phase space of multi-component molecular blends is effective for designing practicable molecular BCs.

## Introduction

Barocaloric (BC) materials are rapidly emerging as leading candidates for enabling commercial solid-state heating and cooling technologies[1–8]. These materials are characterised by their ability to undergo substantial isothermal entropy changes ($|\Delta S_{\mathrm{it}}|$) and adiabatic temperature changes ($\Delta T_{\mathrm{ad}}$) when subjected to hydrostatic pressure. To date, materials exhibiting colossal barocaloric effects ($|\Delta S_{\mathrm{it}}|$ > 100 J K$^{-1}$ kg$^{-1}$) have been discovered across various material families, including plastic crystals[9–17], polymers[18,19] and hybrid organic-inorganic materials[20–23]. As solid-state alternatives, these materials have negligible global warming potential, offering a stark contrast to vapor refrigerants that risk leakage into the atmosphere[24] and can be toxic, flammable and/or greenhouse gases. Furthermore, BC materials have been proposed to achieve coefficients of performance (COP) several times higher than those of existing vapor compression systems[25,26]. This potential for increased sustainability and efficiency is a key driver of BC materials research.

Despite the discovery of numerous materials exhibiting barocaloric effects (BCEs), the search for a technologically relevant BC material continues because of a need to balance multiple practical requirements. Viable BC materials must not only match their phase transition temperatures to the desired application but should, ideally, operate reversibly, with minimal hysteresis and within a pressure limit of ~1 kbar[16,26]. Achieving this without compromising the enthalpy and entropy changes of the phase transition can be challenging and it is often the case that the headline colossal entropy change promised by a new material cannot be accessed in a reversible and practicable device. Amongst the candidate BC materials, PCs are perhaps the most extensively studied, owing to the large entropy changes associated with their order-disorder phase transitions, driven by molecular reorientations. The first colossal BCE was discovered in the PC neopentyl glycol, $(CH_3)_2C(CH_2OH)_2$, (NPG)[9,10], and subsequent studies have shown that related materials, such as pentaglycerine, $(CH_3)C(CH_2OH)_3$, (PG), also display a BCE[11]. However, these materials suffer from significant reversibility issues arising from supercooling effects[27,28] that cause hysteresis in the temperatures at which heat can be rejected and absorbed. Phase transition reversibility can only be achieved at substantially elevated pressures, which is unsuitable for a commercialised device. Additionally, as NPG exhibits a BCE near room temperature, its operational temperature is not ideal for refrigeration purposes[7], and the BCE of PG occurs at temperatures unsuitable for most heating and cooling applications.

Beyond the context of BC applications, forming solid solutions of the neopentyl plastic crystals NPG, PG and pentaerythritol, $C(CH_2OH)_4$, (PE), has been investigated in detail[29–34]. It is well known that the solid-solid (S-S) phase transition temperatures can be tailored through compositional variation. It has also been shown that the doping of pure NPG with organic and inorganic molecules can improve its thermal conductivity[35–38], but without clear improvement of BC performance. The three neopentyl plastic crystal molecules are shown schematically in Fig. 1a, all of which have a tetrahedral structure and differ in the number of hydroxyl units that underpin the hydrogen bonding networks in the ordered crystal (OC) phase. All three PCs undergo S-S phase transitions from a lower entropy OC structure to a higher entropy plastic crystal (PC) structure at temperatures related to the number of intermolecular hydrogen bonds supported by the molecule (~314 K for NPG, ~354 K for PG and ~460 K for PE). Both PG and PE adopt the same OC structure, body-centred tetragonal (BCT, $I\bar{4}$), with hydrogen bonding stabilising a continuous layered structure (Fig. 1c). NPG instead adopts a monoclinic ($P2_1/c$) structure[32], stabilised by intermolecular hydrogen bonds between chains of molecules, resulting in a more needle-like crystal habit[28]. Upon heating, all three materials adopt a face-centred cubic (FCC, $Fm\bar{3}m$) structure in the PC phase.

For practicable BC refrigeration, phase transitions lower than that of NPG (314 K) and with minimal thermal hysteresis are desirable[6,7]. Furthermore, the solid solutions should ideally transition between two well-defined structures, without phase segregation and decomposition concerns, to avoid so-called 'continuous' phase transitions that can span a temperature range of up to 100 K[33]. Fig. 1b shows a schematic of the NPG-PG binary phase diagram reproduced from data calculated elsewhere[32,33]. Here, $\alpha$, $\beta$, and $\gamma$ indicate the BCT, monoclinic and FCC structures, respectively, adopted by the NPG-PG binary solid solution. This phase diagram indicates that the lowest transition temperature of ~300 K (under ambient pressure) that can be reached for a single-phase NPG-PG binary solid solution is for molar compositions of either 8% NPG:92% PG or 60% NPG:40% PG (the latter marked with a red dot in Fig. 1b). Previous calorimetric studies of NPG-PG solid solutions have shown that both compositions exhibit significant thermal hysteresis upon cycling[29] and so are not suitable for practicable devices.

Our hypothesis is that organic dopants could be used to disrupt the hydrogen bonding network that underpins the phase transition, reducing supercooling effects and the problematic thermal hysteresis. We chose small quantities of PE as a dopant, exploiting its similar molecular size and structure but different number of hydroxyl units to the NPG molecule. The 60% NPG:40% PG solid solution is also isostructural with PE, which makes it likely that a single-phase ternary solid solution of the three molecules can be formed. The expected BCT phase, $\alpha$, is characterised by an extensive intermolecular hydrogen bonding network in the $ab$ plane and relatively weaker van der Waals interactions along the $c$-direction[39], as shown in Fig. 1c,d. Our strategy is to introduce different neopentyl molecules into this structure to modify the hydrogen bonding landscape due to the differing numbers of hydroxyl functional groups, ultimately affecting the phase transition conditions for the solid solution. Although binary and ternary solid solutions of the neopentyl PCs have been studied extensively, their BC performance has not been investigated. Furthermore, all our measurements were carried out over a period of approximately 15 months, demonstrating the long-term thermodynamic stability of the solid solutions. This strategy has the potential to optimise BC properties through compositional engineering in neopentyl PCs and other BC molecular systems, such as functionalised adamantane[12,14,40–42], which unlike pure adamantane[43], have transition temperatures close to room temperature.

We present results for a binary solid solution of 60% NPG and 40% PG in comparison to the same composition doped with a small amount of PE, i.e. a ternary solid solution of 60% NPG, 38% PG, and 2% PE and benchmark these materials against pure NPG. Both solutions retain the colossal barocaloric effect observed in NPG and the main result is that we are able to access a greater fraction of that colossal entropy change under reversible, isothermal conditions with pressures ≤1 kbar. In comparison to NPG, the ternary solid solution has a seven times greater reversible isothermal entropy change ($|\Delta S_{\mathrm{it,rev}}|$ = 13.4 J kg$^{-1}$ K$^{-1}$) and its operational temperature span ($\Delta T_{\mathrm{span}}$ = 18 K) is increased by a factor of 20 at the technological pressure threshold of 1 kbar, corresponding to a more than 70 times increase in reversible refrigeration capacity at this pressure. Critically (and in line with previous studies), we can tune the transition temperature of the solid solution to be closer to the operational requirements of cooling applications. To understand these enhancements, we use powder synchrotron x-ray diffraction and quasielastic neutron scattering experiments, which reveal structural and dynamical changes, aiding our understanding of how disrupting the hydrogen bond network can improve the reversibility of the OC→PC phase transition. Neutron spectroscopy fixed window scan techniques reveal that the activation energies for the rotational modes underpinning the BCE are reduced by up to 50% compared to NPG. We propose that the improved reversibility and operational temperature span are facilitated by dopant-induced heterogeneous nucleation sites and a large ($\alpha$,$\gamma$) phase co-existence region spanning around 30 K during thermal cycling. These properties arise from a disruption to the hydrogen bond network caused by substitutional doping. Our findings pave the way for rational design in creating new BC PCs, accessing significant compositional phase spaces and enabling tailored BC performance.

# Results

### Comparison of barocaloric effect using variable-pressure calorimetry

Full characterisation was only conducted on the most promising compositions. We therefore focused on the binary NPG:PG composition that was predicted to have the lowest isothermal phase transition temperature under applied pressure and that was also expected to transition between two well-defined phases (see Fig. 1): the binary solid solution of 60% NPG and 40% PG (herein denoted as NPG-PG). Ambient-pressure calorimetry (Supplementary Fig. 1) was used to determine the optimal ternary composition, a solid solution of 60% NPG, 38% PG, and 2% PE (herein denoted as NPG-PG-PE). Calorimetric measurements under variable applied pressure were then conducted on both solutions in comparison with NPG, as shown in Fig. 2. In each case, the latent heat is attributed to the reversible OC→PC phase transition. The comparison is evaluated quasi-directly from isobaric calorimetric peaks (Supplementary Fig. 2a,b), using calculations outlined in the Methods Section. This standard method for determining reversible BC entropy changes[11,44] allows a conversion from data collected during isobaric thermal cycling to approximate data relevant to pressure cycling. The former is experimentally efficient whilst the latter represents the operation of a realistic BC device. In these plots, the OC→PC (PC→OC) phase transition corresponds to the peaks with positive(negative) $\Delta S_{\mathrm{it}}$ that would be driven by decompression(compression) sweeps in a real device. In each case, the maximum value of $\Delta S_{\mathrm{it}}$ is achieved at the corresponding peak temperature and so our aim was to minimise the hysteresis. This would bring the peaks into alignment for compression and decompression steps and thereby maximise the absolute reversible entropy change ($|\Delta S_{\mathrm{it,rev}}|$) at a fixed temperature.

A comparison of Fig. 2a (NPG-PG) with Fig. 2b (NPG-PG-PE) reveals that both materials exhibit 'colossal' barocaloric effects with $|\Delta S_{\mathrm{it}}|$ ~ 200 J K$^{-1}$ kg$^{-1}$. Critically, NPG-PG-PE displays significantly reduced hysteresis in the first-order phase transition, resulting in an enhancement of its reversible BC performance compared to NPG-PG. In contrast, NPG-PG, shown in Fig. 2a, has no reversible entropy change at any of the pressures measured here ($|\Delta S_{\mathrm{it,rev}}|$ = 0), as the peaks for PC→OC and OC→ PC do not overlap at any temperature. Similarly, the reversible entropy change for NPG is known to be small when driven with pressures around 1 kbar[9,10]. In both cases, the implication is that a compression/decompression cycle under the isothermal conditions of a BC refrigerator based on NPG or NPG-PG could not access both the OC→PC and PC→OC phase transitions when operated with 1 kbar.

NPG-PG-PE has a large region of reversible entropy change, with overlapping PC→OC and OC→ PC curves around 300 K. Although the OC→PC peak phase transition temperature is nearly identical for both materials at all applied pressures, the improvement arises because the supercooled PC→OC transition occurs at higher temperatures for NPG-PG-PE (Fig. 2b) and has a distinct asymmetric character. Furthermore, both OC→PC and PC→OC phase transitions are broader for NPG-PG-PE, and are especially pronounced during the supercooled PC→OC transition, when compared to NPG-PG. In Fig. 2c, a comparison of NPG, NPG-PG and NPG-PG-PE with an applied pressure of $|\Delta p|$ = 1000 bar shows that NPG-PG-PE exhibits the largest reversible temperature span ($\Delta T_{\mathrm{span}}$), as indicated by the shaded blue region. NPG-PG has no reversible isothermal entropy change at this pressure ($|\Delta S_{\mathrm{it,rev}}|$ = 0, $\Delta T_{\mathrm{span}}$ = 0), highlighting the significant improvement of the BC performance from the addition of just 2% PE to the binary solid solution. Figure 2d demonstrates that at the critical technological threshold pressure of $|\Delta p|$ = 1000 bar, NPG-PG-PE produces a maximum reversible isothermal entropy change of up to $|\Delta S_{\mathrm{it,rev}}|$ ~ 13.4 J $K^{-1}$ $kg^{-1}$ over a temperature span of $\Delta T_{\mathrm{span}}$ ~18.0 K. In comparison, pure NPG under the same pressure change produces a ~90% lower reversible isothermal entropy change of $|\Delta S_{\mathrm{it,rev}}|$ ~ 1.9 J $K^{-1}$ $kg^{-1}$, with a narrow temperature span of $\Delta T_{\mathrm{span}}$ ~ 0.9 K at a temperature of ~314 K. This corresponds to a reversible refrigerant capacity ($\mathrm{RC_{rev}}$) of 73 J $kg^{-1}$ for NPG-PG-PE and 1 J $kg^{-1}$ for NPG at 1 kbar (see Supplementary Fig. 2d for $\mathrm{RC_{rev}}$ values at other pressures). This improvement of BC performance at 1 kbar highlights the capabilities of NPG-PG-PE relative to pure NPG. Further examination of Fig. 2b,c reveals that the broader regime of reversible $|\Delta S_{\mathrm{it}}|$ in NPG-PG-PE results from the increase in asymmetry of the PC→OC phase transition as compared to NPG-PG. The asymmetry in the phase transitions is marked by an extended tail around 300 K in Fig. 2c, clearly contributing to enhanced reversible BC performance. Notably, while the hysteresis in NPG-PG-PE is reduced, the sensitivity of the transition temperature to pressure ($\frac{dT_0}{dp}$) remains similar in both NPG-PG and NPG-PG-PE (Supplementary Fig. 3 and Supplementary Fig. 4).

**Nucleation density from infra-red imaging**

Infra-red (IR) imaging of BC samples provides invaluable spatially-resolved insight into heat emission and absorption, complementing the calorimetry analysis. We imaged all three samples during cooling using an IR camera (Supplementary Fig. 5-7), with the main observations summarised in Fig. 3. As the samples transition from PC to OC phases, the heat released is observed as a local increase in temperature that is readily observed in differences between successive frames of a sequence of infrared images. An advantage of this approach is that it highlights just material that is undergoing the phase transition, either at a nucleation point or at a propagation front; another advantage is that it is more directly relatable to the differential calorimetry measurements discussed above.

Pure NPG (Supplementary Fig. 5) shows a small number of nucleation events followed by the rapid propagation of phase transition wave fronts through the material, as observed previously[28,45]. For NPG, the phase transition is clearly linked to a crystallisation process, with the phase transition advancing along needle-like domains. Neither NPG-PG-PE nor NPG-PG (Fig. 3, Supplementary Fig. 6 and Supplementary Fig. 7) demonstrate the same behaviour. As they become super-cooled, both samples are characterised by a greater number of more localised nucleation events that appear and are quenched without significant propagation of transition fronts through the material. There is no such directionality to the propagation, which suggests a more polycrystalline, isotropic underlying structure with smaller grains and a greater mosaic spread. The direct comparison between NPG-PG and NPG-PG-PE seen in Fig. 3 suggests that the addition of PE dramatically increases the nucleation density in NPG-PG-PE compared to NPG-PG. This larger number of more discrete nucleation events may explain the protracted nature of the PC→OC phase transition measured in calorimetry, which will be discussed further below. Videos showing the IR imaging through the phase transition for each sample can be found in Supplementary Movies 1-3.

**Phase transition-induced structural changes**

Previous studies have shown that solid solutions of neopentyl-based PCs exhibit a variety of crystal structures[33,34,39]. For the composition indicated by the red dot in Fig. 1b, we expected NPG-PG to adopt the same OC BCT structure as PG[30,33], which has an extensive hydrogen bond network in the $ab$ plane and van der Waals interactions between layers in the $c$-direction (Fig. 1c,d). The lattice symmetry of NPG-PG-PE, NPG-PG and PG was compared using powder synchrotron x-ray diffraction (XRD) during heating and cooling under ambient pressure (Supplementary Fig. 8-10). Analysis of the NPG-PG-PE structure is presented in Fig. 4, with similar datasets for the other two samples presented in Supplementary Fig. 11 and Supplementary Fig. 12. (Note that comparison with NPG was not included as it adopts a different, monoclinic structure in its low temperature phase.) All three samples show a thermally-activated expansion from a single-phase BCT structure to a single-phase FCC structure: we confirm that NPG-PG adopts the structure predicted by Fig. 1b and also confirm that the addition of 2% of PE does not alter that crystal symmetry. The low temperature structure, including molecular conformations, was confirmed using single crystal electron diffraction (Supplementary Information Note 1) for both PG and NPG-PG-PE at 150 K, which verified that they are isostructural, with varying occupation of hydroxymethyl and methyl groups in accordance with their composition.

A comparison of the refined lattice parameters (also plotted in Fig. 5a) shows that NPG-PG-PE is ~0.4% contracted in the $ab$ plane and ~1.2% expanded along the $c$-axis compared to PG. A similar trend is observed in comparing NPG-PG with PG, and there is a slight expansion of the $ab$ lattice with the addition of PE to form NPG-PG-PE. Such subtle anisotropic changes arise from a combination of steric effects and the potential for additional hydrogen bonding. Addition of hydroxyls will increase the physical size of molecules, potentially expanding the surrounding lattice and consequently reducing the strength of hydrogen bonding. On the other hand, if the hydroxyls are oriented favourably, then increased hydrogen bonding may reduce the local intermolecular separation. Although the hydroxyls can orient to lie within the $ab$ plane, recent simulations also suggest that the degree of hydrogen bonding may be less than expected on the basis the hydroxyl count [46]. In the present case, expansions in the $ab$ plane with increased hydroxylation suggest that steric effects and hydroxyl orientation are important.

For all three samples, individual phases show approximately linear thermal expansion away from the phase transitions. The OC→PC phase transition is then characterised by a substantial increase in intermolecular separation in the $ab$ plane and corresponding reduction of the intermolecular spacing along the $c$-axis (Fig. 4a,b), producing an overall volumetric expansion (Fig. 4c). Overall, a volume change of $\Delta V$ = 4.8% (Fig. 4c) occurs through the phase transition in NPG-PG-PE, obtained from the refinements (Supplementary Fig. 10) at 292 K (BCT) and 319 K (FCC). It is notable that that this volume change is close to that of pure NPG[11]. These observations are consistent with the loss of hydrogen bonding in the $ab$ plane as the molecules start to rotate, transitioning to an FCC structure with non-directional bonding. The equilibrium phase transition temperatures (on heating, $T_0$) are $T_0$ = 307 K for NPG-PG-PE, $T_0$ = 302 K for NPG-PG and $T_0$ = 354 K for PG, which compare well with the ambient pressure calorimetry data (Supplementary Fig. 1) and demonstrate an ability to reduce the transition temperature of PG through appropriate blending and doping.

Closer analysis of XRD data reveals that the NPG-PG-PE phase transition occurs over a wider temperature range than that of isostructural PG. Figure 5a compares the temperature dependence of the BCT lattice parameters for NPG-PG-PE and PG after four initial thermal cycles (lattice parameters shown in Supplementary Fig. 13) as they are heated towards their transition temperatures (NPG-PG is shown in Supplementary Fig. 14 to behave similarly to NPG-PG-PE). The shaded areas indicate the temperature ranges where both OC and PC phases co-exist, determined from the refined phase fraction in Fig. 5b. For both NPG-PG (Supplementary Fig. 11) and NPG-PG-PE the $\alpha$ and $\gamma$ phases coexist across a range of ~30 K (pink shading), substantially wider than the ~8 K span for PG (grey shading). For $\frac{T}{T_0}$ < 0.96 the temperature dependence of the unit cell parameters of the BCT phase in all materials is dominated by the roughly linear trends expected for thermal expansion. The main contribution to the volume expansion of the BCT phase is the relatively large changes in $c$ (note the scales in Fig. 5a), presumably due to the weaker van der Waals bonding in this direction. Approaching the transition ($\frac{T}{T_0}$ > 0.96), there is a notable, non-linear increase in the expansion of the NPG-PG-PE $a$ lattice parameter (blue, left axis). This trend is not observed in PG, where a linear trend continues up to the transition. The non-linear behaviour in both NPG-PG and NPG-PG-PE (compared with PG in Supplementary Fig. 14) systems suggests disruption of the hydrogen bond network beyond simple thermal expansion as $\frac{T}{T_0}$ = 1 is approached.

A similar anisotropy is observed in the breadth of $hkl$ diffraction peaks close to the transition temperature. We fitted Stephens' anisotropic broadening parameters,[47] $S_{hkl}$, to $hkl$ reflections from each sample (see Methods Section and Supplementary Fig. 15). Figure 5b compares the temperature dependence of $S_{400}$ for PG and NPG-PG-PE, showing that $h00$ type reflections begin to broaden in NPG-PG (Supplementary Fig. 15) and NPG-PG-PE systems for $\frac{T}{T_0}$ > 0.96. In contrast, increased broadening of the PG $h00$ type reflections are only observed at the start of the full phase transition around $\frac{T}{T_0}$~1.00. $S_{004}$ fits to the broadening of $00l$ type reflections show relatively smaller increases compared to $S_{400}$ at the phase transition for all materials, whilst $S_{220}$ type broadening is only required to fit the data for NPG-PG and NPG-PG-PE (Supplementary Fig. 15). These observations again highlight a clear anisotropy in the thermal expansion of the PCs close to their phase transition, and also clear differences between PG and the blends NPG-PG-PE and NPG-PG. Around $\frac{T}{T_0}$ ~ 1.00, the first-order region of the phase transition in all samples is characterised by a rapid increase in $S_{400}$ that correlates with a rapid change in the phase fraction. However, the more gradual variations in lattice parameter and peak broadening observed for both solid solutions suggest a weakening and perhaps localised disruption of the hydrogen bond network that define what we term the weakly first-order region of the transition. Similar protracted phase transitions are observed upon cooling from the PC back to the OC phase. Supplementary Fig. 16 includes the $S_{hkl}$ parameters on cooling for all three samples, where NPG-PG-PE has significantly enhanced peak broadening compared to the other two materials, suggesting a greater degree of structural disorder for this sample. As discussed below, we propose that this increased disorder, particularly in the supercooled state, is intricately linked to the increased nucleation events observed by IR imaging during the supercooled transition in NPG-PG-PE.

**Molecular dynamics from quasielastic Neutron Scattering**

Neutron scattering techniques allow dynamical changes in NPG-PG-PE to be correlated with the observed thermal and structural behaviour, as previously shown for other BC PCs[10,43,48–50]. Quasielastic neutron scattering (QENS) probes the molecular dynamics of hydrogen-containing materials by measuring low energy transfers, typically ranging from meV to μeV, which correspond to dynamic timescales of picoseconds to nanoseconds. QENS datasets contain intensity as a function of energy ($E$) transfer and momentum ($Q$) transfer ($S(Q,E)$) and feature an elastic peak (centred at $E$ = 0) that is broadened by quasielastic scattering caused by rotational and translational diffusion processes within the sample[51]. Thus, the degree of broadening is directly correlated with the dynamics occurring within the material. In recent work, we used QENS to identify the key molecular rotational modes responsible for the BCE in NPG and employed inelastic fixed-window scans (FWS) to directly observe a hysteresis in the molecular dynamics[50]. Here, we again used the IN16B spectrometer at the Institut Laue-Langevin (ILL), which uniquely enables such inelastic FWS measurements by capturing the quasielastic signal within specific narrow energy ranges. Importantly, this technique allows molecular dynamics to be tracked whilst the sample undergoes thermal cycling, providing direct comparison with diffraction and calorimetry measurements. Given the structural similarity between NPG, PE and PG, differing only in the ratio of hydroxymethyl to methyl groups, we apply similar QENS/FWS data analysis to that used for NPG to understand the molecular dynamics in NPG-PG-PE. Given that the binary NPG-PG did not exhibit a reversible BC effect at 1.2 kbar or below, we did not perform QENS analysis on this material. Details regarding the QENS/FWS data acquisition and analysis can be found in the Methods Section and Supplementary Information Note 3.

Figure 6a shows a thermal cycle for NPG-PG-PE using ambient pressure calorimetry with a scanning rate of 1 K min$^{-1}$. Thermal hysteresis is observed in the difference between the heating and cooling peaks, and the data in Fig. 2b show that the peak overlap, and hence the accessible latent heat under isothermal conditions, increases with pressure. In Fig. 6a there is a temperature range of approximately 10 K where the evolution of latent heat is reversible, even at ambient pressure, which is atypical for PCs. An ambient pressure equilibrium phase transition temperature of $T_0$ = 307 K, in good agreement with the XRD data, is determined from the peak temperature on the heating step and a supercooled phase transition temperature of $T_{\mathrm{SC}}$ = 283 K is obtained from the intersection of the cooling peak inflection point gradient with the temperature axis. Both temperatures are indicated in Fig. 6a with black arrows. The onset is used for the cooling transition due to significant asymmetry, characteristic of supercooled phase transitions. The large phase coexistence region obtained from the XRD phase fraction data on heating is indicated by the double-headed arrow in Fig. 6a, demonstrating that calorimetry and XRD measurements are well correlated. The $\Delta T$ of 30 K, shown in both techniques, is an order of magnitude greater than that of NPG obtained at the same ramp rate[28]. This implies a gradual, broadened first-order phase transition where both phases coexist over a broad temperature range. It also indicates that domain kinetics may play an important role in the phase transitions of NPG-PG-PE. The calorimetry data can be directly correlated with inelastic FWS (IFWS) measurements, also obtained at ambient pressure and a heating rate of 1 K $\mathrm{min}^{-1}$, shown in Fig. 6b. For IFWS measurements, rather than measuring quasielastic scattering across all energies at a constant temperature, as is usual in QENS ($S(Q,E)$), instead a specific energy window is chosen to measure quasielastic scattering as a function of temperature ($S(Q,T)$). This makes it possible to infer contributions from modes in that energy window and to track their temperature dependence. A larger intensity in the IFWS plot implies increased quasielastic scattering within the chosen energy windows, in this case located at $\pm$ 3 μeV. Thus, a sudden increase(decrease) in IFWS intensity on heating(cooling), as indicated by $T_0(T_{\mathrm{SC}})$ in Fig. 6b, corresponds to increased(decreased) molecular dynamics in the sample. This is expected when observing the liberation(freezing) of the PC modes at $T_0(T_{\mathrm{SC}})$. Thermal hysteresis in the molecular dynamics is observed as a retention of the quasielastic scattering signal to lower temperatures. The parameters $T_0$, $T_{\mathrm{SC}}$ and $\Delta T$ obtained from ambient pressure calorimetry and XRD are well aligned with the features observed in the IFWS data, indicating that it is possible to correlate the thermal, structural and dynamical information of NPG-PG-PE.

To characterise the dynamic modes, full QENS scans were obtained on heating by isothermally measuring at 200 K, 230 K, 260 K, 330 K and 360 K with the scans either side of $T_0$ shown by the yellow stars in Fig. 6b. The fitting of QENS scans at 260 K and 330 K for $Q$ = 0.83 Å$^{-1}$, representative of other $Q$ values (Supplementary Fig. 19), can be seen in Fig. 6c,d where the red dashed lines indicate the energy windows ($\pm$3 μeV) at which the IFWS in Fig. 6b were obtained. The fitting was to the global dataset consisting of both QENS and IFWS, rather than for individual spectra, and was constrained using the IFWS data across all $Q$ values (Supplementary Information Note 3). Within this model, each Lorentzian ($L_{\mathrm{i}}$) represents a dynamical mode detected within the sample, with a characteristic amplitude and linewidth frequency. The full width at half maximum (FWHM) linewidth of the fitted Lorentzians ($L_{\mathrm{i}}$) as a function of $Q$, $\Gamma(Q)$, is shown in Fig. 6e. Here, the $Q$-independent behaviour indicates localised rotational dynamics rather than long range diffusion processes. As was also observed in pure NPG previously, we observe one mode below the phase transition (260 K) and two additional modes in the PC phase above 330 K. We assign these modes as $L_1$ = methyl rotation with radius $r_1$, $L_2$ = hydroxymethyl rotation with radius $r_2$ and $L_3$ = molecular reorientation with radius $r_3$, through calculating their elastic incoherent structure factors ($\mathrm{EISF}(Q)$, Fig. 6f) and fitting the data to known geometric models with good agreement (see Supplementary Information Note 3 and Supplementary Fig. 20 for the geometric origins of $r_{\mathrm{i}}$). The relative magnitudes of the fitted radii are consistent with the modes that they represent, i.e. $r_1$ (methyl rotation) < $r_2$ (hydroxymethyl rotation) < $r_3$ (molecular reorientation). We included a fractional fitting parameter ($f$) to account for scatterers that do not participate in the respective process [49,50,52–54]. For Mode 1, the methyl rotation radius was fixed to the value obtained from crystallography[55] and we find that $f$ is lower than it is for pure NPG[50], which is consistent with NPG-PG-PE having ~ 20% fewer methyl groups. In fact, for NPG-PG-PE the fraction of $^1$H atoms in methyl and hydroxymethyl groups is 0.395 and 0.605, respectively, agreeing very well with the fitted $f$ parameters of 0.37 for Mode 1 and 0.60 for Mode 2. The fit for Mode 3, molecular reorientation, compares well with that obtained for pure NPG[50], where a fractional fitting parameter $f$ < 1.0 is also observed. Since all $^1$H atoms are expected to be involved in this rotational motion, $f$ should be unity, and a possible explanation is that not all orientations are accessible at the phase transition temperature[48,50,56]. Together, these fits indicate that the rotational modes present in NPG-PG-PE are very similar to those of pure NPG, which is to be expected given the chemical similarity of the neopentyl plastic crystals.

We now compare features in the calorimetry data in Fig. 6a with those in the IFWSs in Fig. 6b. On heating, the increase in IFWS signal matches the endothermic latent heat peak in calorimetry, with the inflection point of the IFWS signal approximately coinciding with the peak latent heat temperature. These observations correspond to the OC→PC phase transition. Since the IFWS signal increases during this transition, it is evident that additional dynamical modes are activated on heating. On cooling, the calorimetry data appear to show two-component behaviour in the exothermic latent heat peak. Firstly, the latent heat peak slowly increases from 300 K to 288 K, indicating the weakly first-order region, followed by a sharp increase centred around 280 K, characteristic of a supercooled phase transition. The slow decrease of IFWS intensity with cooling indicates 1) a gradual slowing of molecular dynamics and/or 2) a reduction in the number of scatterers exhibiting the PC molecular dynamics. This weakly first-order region is followed by a sharp decrease in the IFWS intensity, coinciding with the sharp exothermic latent heat peak in Fig. 6a. This indicates a rapid decrease in the detected molecular dynamics. The PC→OC phase transition (i.e. on cooling) is characterised by the formation of the hydrogen bond network present in the BCT phase, and thus it is expected that the low IFWS signal below ~ 280 K is due to PC dynamics being frozen by the existence of an extensive hydrogen bond network that immobilises the rotation of the molecules and their hydroxymethyl groups. Figure 7a shows the IFWS data measured at $\pm 3$ μeV on cooling from 366 K to 100 K for $Q$ = 0.83 $Å^{-1}$ at a rate of 1 K $min^{-1}$, directly comparable to our calorimetry and XRD measurements. Further fits for low- and high-$Q$ IFWS data can be found in Supplementary Fig. 21. Figure 7b shows the corresponding IFWS data on heating from 240 K to 355 K. We will first discuss the cooling data as it contains a larger temperature range. The peak centred around 200 K in Fig. 7a corresponds to Mode 1, the methyl rotation, as we observed previously in NPG[50]. Peaks in the IFWS data are attributed to the Lorentzian terms 'moving through' the fixed energy window as the associated mode changes in frequency with changing temperature (broadening into the background for heating and narrowing into the elastic line for cooling). The inset plot compares the peak intensity of NPG-PG-PE with that of NPG after normalisation to the low-$T$ region, showing that the NPG-PG-PE peak is less intense than in NPG. This reduced intensity can be attributed to the solid solution containing 21% fewer methyl groups than NPG, which is consistent with the measured intensity ratio of 0.76 ± 0.02. Fitting the thermal variation of Mode 1 (methyl rotation, blue solid line) as described previously[57], we model a temperature dependence by assuming an Arrhenius law for the linewidth ($\Gamma(Q)$):

$$\Gamma(Q) = \Gamma_0(Q)e^{-\frac{E_a}{k_B T}}. \quad (1)$$

By restricting fitting of the data to $T$ < 260 K we ensure that the phase transition region is not included in the fit, and we calculate an activation energy ($E_a$) of 94 ± 1 meV for methyl rotation in the OC phase. The value obtained is close to the activation energy of 100 meV obtained for pure NPG[50], demonstrating the consistency of our model. The sharp intensity change observed upon cooling from approximately 300 K to 290 K is attributed to the PC→OC phase transition, during which the rotational modes present in the PC phase become immobile as the extensive hydrogen bond network of the OC phase forms. The peak centred around 340 K contains the intensity contribution from both Mode 2, hydroxymethyl rotation, and Mode 3, molecular reorientation. It is possible to separate the contributions of each mode to the total IFWS intensity in the PC phase. This was achieved by globally fitting all obtained IFWS cooling data with the QENS data at 330 K and 360 K to constrain the intensity ratio of Mode 2 and Mode 3 determined by the QENS fits (see Supplementary Information Note 3 for further details). Furthermore, it is only meaningful to fit the IFWS data in regions where the material is fully in the PC phase. To determine the range of IFWS data to fit we parameterise the FCC phase fraction ($\Phi_{FCC}$) by fitting error functions (Supplementary Information Note 2) to the full heating and cooling synchrotron XRD data (Fig. 4).

By fitting the IFWS cooling data for temperatures where $\Phi_{FCC}$ > 0.98, shown by the grey shaded region in Fig. 7a, we obtain $E_a$ for Mode 2 (hydroxymethyl rotation, green dashed line) and Mode 3 (molecular reorientation, orange dashed line) of 460 $\pm$ 5 meV and 300 $\pm$ 3 meV, respectively. These values are both approximately 50% lower than the corresponding activation energies of the same modes observed experimentally in NPG [50]. Furthermore, the measured activation energy of hydroxymethyl rotation in NPG-PG-PE is close to that of two hydrogen bonds (~500 meV[48]), suggesting that the energy barrier to hydroxymethyl rotation is highly correlated with the hydrogen bond network in the ternary solid solution. We note that obtaining precise activation energies of multiple overlapping modes is not trivial and is enabled by our ability to constrain the mode intensity ratios by including the QENS datasets in the global fitting procedure. The reduction in activation energies of Modes 2 and 3 agree with the observations from the XRD data that the energy landscape of the hydrogen bond network in NPG-PG-PE is significantly altered as compared to NPG. Curiously, the activation energy for the hydroxymethyl rotation is found to be larger than that for the molecular reorientation, as has also been observed experimentally[50] and theoretically[48] for NPG. Further support for this interpretation is gained from an inspection of the relative magnitudes of the fitted pre-exponential factors ($\Gamma_0(Q)$) used in the Arrhenius law in Eq. 1. The full $\Gamma_0(Q)$ data can be found in Supplementary Fig. 22. Converting the pre-exponential factors to attempt frequencies yields ~140 THz and ~100 GHz for hydroxymethyl rotation and molecular reorientation, respectively. These values are consistent with molecular rotational attempt frequencies around the microwave region [58]. They demonstrate that although the hydroxymethyl rotation has a higher energy barrier to overcome, it also has an attempt frequency around 3 orders of magnitude greater than that of molecular reorientation, which may reconcile the relative magnitudes of their activation energies. Together, these observations suggest that hydroxymethyl rotation is concurrent with the extinction of the hydrogen bond network, and unlocks whole molecular rotation, with its lower activation energy. Therefore, lowering the energy barrier for hydroxymethyl rotation enables access to the large entropy change associated with molecular reorientations at lower temperatures.

Thermodynamically, we do not expect a difference in activation energies or attempt frequencies for cooling versus heating, only a difference in the transition temperature and phase coexistence region. By fixing $E_a$ and $\Gamma_0(Q)$ obtained from the fits of the IFWS cooling data, we can assess how well the derived model is able to describe the heating data (Fig. 7b). The purple dashed lines in Fig. 7a,b show the FCC phase fraction ($\Phi_{\mathrm{FCC}}$) on cooling and heating as determined from the XRD data (Supplementary Information Note 2), respectively, with $T_{\mathrm{SC}}$ and $T_0$ indicated by the vertical black arrows. Here we present a method to utilise $\Phi_{\mathrm{FCC}}$ to fit the total temperature-dependant quasielastic behaviour ($S(T)$, given by the red dotted line) through the phase transition for 0.02 < $\Phi_{\mathrm{FCC}}$ < 0.98. To do this we weighted the previously obtained Lorentzians ($L_1$, $L_2$ and $L_3$) from fitting the IFWS cooling data by the corresponding phase fraction according to:

$$S(T) = (1 - \Phi_{\mathrm{FCC}})L_1 + \Phi_{\mathrm{FCC}}(L_2 + L_3). \qquad (2)$$

As can be seen by the trends in Fig. 7a,b the red dotted lines that describe $S(T)$ provide good fits for the phase transition regions on both cooling and heating. This highlights the need to consider modes from both the OC and PC phases throughout the broad phase coexistence regions. Note that Mode 1 (methyl rotation) is also active in the PC phase, however the FWHM linewidth of this mode (~500 μeV[50]) is close to that of IN16B's dynamic range (560 μeV), and as such we are not directly sensitive to it. The ability to combine XRD, QENS and FWS data in this way provides a powerful tool for describing dynamical mode behaviour even during phase transitions, where phase heterogeneity is significant.

## Discussion

The observation that BC properties can be tuned intrinsically is exemplified in Fig. 2. Choosing an NPG:PG ratio of around 60:40 results in a single-phase material that exhibits a $T_0$ that is 10 K lower than NPG and 50 K lower than PG, around the regime required for BC cooling. The main result of this study is that adding a small amount of a third component, in this case PE, leaves $T_0$ largely unchanged but drastically improves the reversibility of the BCE as exemplified in Fig. 2d. The improvement is due to an asymmetric increase in the $\alpha, \gamma$ phase-coexistence region of the PC→OC phase transition, which results in a significant overlap of latent heat in the hysteresis region, as shown in both Fig. 2c and Fig. 6a. Furthermore, the solid solutions of NPG-PG and NPG-PG-PE exhibit enhanced phase coexistence regions, which is more pronounced at higher pressures, as shown in Fig. 2a. This explains the broad temperature range, $\Delta T_{\mathrm{span}}$ ~18.0 K, over which NPG-PG-PE exhibits a reversible entropy change. This flexibility provides potential opportunities for multi-component cascading barocaloric devices[59], where the composition of the solid solution, and hence the transition temperature, is engineered to vary within a heating/cooling device to improve efficiency.

Our results also reveal the structural changes occurring within the phase coexistence region and the nature of the doping in NPG-PG-PE. The synchrotron XRD data presented in Fig. 5 demonstrates that the OC→PC phase transition exhibits a broadened first-order behaviour within the temperature range $0.96 < \frac{T}{T_0} < 0.99$, characterised by tracking the anisotropic diffraction broadening parameters $S_{400}$ and $S_{220}$, indicative of weakened hydrogen bonding in the $ab$ plane. This contrasts with the sharply first-order behaviour of isostructural PG, which shows negligible $S_{220}$ broadening and a sharp transition of $S_{400}$ increasing within a narrow temperature range around $T_0$ (Supplementary Fig. 15). These observations are also supported by the IFWS data presented in Fig. 7, where we calculate activation energies of the PC modes that are approximately 50% lower than those of NPG. In NPG-PG-PE, the change in quasielastic signal is much more gradual through the phase transition (on both heating and cooling) than pure NPG (Supplementary Fig. 23), due to the significant mixing of OC and PC rotational modes within the phase coexistence region.

The improvement of BC performance in NPG-PG-PE compared to NPG-PG comes primarily from a reduction in thermal hysteresis, which is caused by a shift of the PC→OC phase transition to higher temperatures, as exemplified in the entropy change data in Fig. 2c. The diffraction results (Fig. 5 and Supplementary Fig. 15) indicate that the binary and ternary have increased short-range disorder or strain over a broad temperature range when compared to PG. Furthermore, the ternary exhibits significantly higher strain than the other materials, particularly on cooling (Supplementary Fig. 16). In turn, this suggests that dopant PE molecules, even at 2%, induce further significant local distortions in the host NPG-PG structure. It is tempting to conclude that the local, often transient hotspots observed in IR while cooling the NPG-PG-PE sample correspond to inclusions of higher PE concentration that undergo the PC→OC transition before the surrounding matrix but are insufficient to trigger a wave of transition throughout the material. A critical question is therefore the location of the PE dopant molecules, which may help explain why such a small quantity of PE has such a profound effect on the phase-change behaviour. It is conceivable that the PE molecules could be incorporated into the BCT structure interstitially. However, our XRD data supports direct substitution of NPG molecules for PE. In accordance with our initial hypothesis, direct substitution is facilitated by the structural similarities between the neopentane molecules.

The substitutional model is further evidenced by our single crystal electron diffraction measurements (Supplementary Information Note 1, Supplementary Fig. 17 and Supplementary Fig. 18), where fractional occupation of hydroxymethyl and methyl groups in the refined structures are in line with the composition of NPG-PG-PE, assuming a substitutional inclusion of NPG and PE (as shown in Supplementary Tables S1-S6). On a substitutional site, PE is directly involved in the hydrogen bond network and so would be expected to affect bonding within the $ab$ plane, and hence explain the enhanced broadening of $hk0$ type reflections in NPG-PG-PE compared to NPG-PG and PG. It may simply be localised strain around the substitutional PE molecules that facilitates localised phase changes and thereby extends the phase coexistence region, specifically on cooling.

Our strategy has been to employ substitutional organic doping of molecular crystals with similar molecules as a general method of improving their thermal and pressure reversibility. Looking forward, this strategy will also be applicable to, but is not limited to, the adamantane PC family, where functional substitution of H atoms with more electronegative groups such as -Cl and -Br results in materials that exhibit a colossal BCE at distinct temperatures[12,14,46]. These functional substitutions do not significantly change the structure of the PC phase but may drastically alter the orientational molecular dynamic processes[41,43], OC structure[60] and barocaloric performance[46]. As such, the adamantane PC family may be suitable for BC enhancement through small compositional doping of organic derivatives.

## Conclusions

We have demonstrated a strategy to improve plastic crystal barocaloric performance with respect to operating temperature ($T_0$), reversible temperature span ($\Delta T_{\rm span}$) and reversible entropy change ($|\Delta S_{\rm it}|$). Using NPG as an exemplar, and by forming the solid solution of NPG-PG-PE in a molecular ratio of 60:38:2, we reduce $T_0$ by 10 K and increase $\mathrm{RC}_{\rm rev}$ by a factor of seventy, arising from the increase in $\Delta T_{\rm span}$ by a factor of twenty and increase $|\Delta S_{\rm it}|$ by a factor of seven at 1 kbar. This improvement comes from an enduring OC-PC phase coexistence that broadens the phase transition between low- and high-entropy phases at both ambient and elevated pressures. We present structural and dynamical evidence for a broad phase coexistence region that is characterised by a disrupted hydrogen bond network. We propose that disrupting the network through substitutional doping is a strategy for improving the reversibility of the barocaloric phase transition. Our study showcases a new strategy for designing practicable molecular crystal barocalorics with tuneable operating temperatures and reduced hysteresis. Given the vast compositional phase space for molecular crystal solid solutions, the prospects for applying this strategy to other molecular systems are exciting.

## Methods

### Sample preparation

Powder samples of NPG, PG and PE (99% purity, Sigma-Aldrich) were used as received. Powders were mixed and then dissolved in ethanol before being left to recrystallize under ambient conditions for a period of 1 week. Each sample was then subjected to thermal treatment consisting of heating to 350 K for 30 minutes and cooling to 265 K for 24 hours, which was to ensure that the samples were in their low temperature phase. Samples were then stored in sealed containers at room temperature. Some of the NPG-PG-PE sample was stored in the freezer at 265 K for three months and its $T_0$ value, obtained from ambient pressure calorimetry, was compared to the sample stored at room temperature to confirm that no phase segregation had occurred.

### Calorimetry

All calorimetry measurements and analysis reported here have been performed on samples that have been thermally cycled around the OC→PC transition at least 10 times prior to remove virgin effect variations[28]. Supplementary Fig. 2c demonstrates the stability of the calorimetric behaviour of NPG-PG-PE after the first thermal cycle. Calorimetric measurements under ambient pressure were obtained using a TA instruments DSC 250 at a temperature-ramping rate of $\frac{|dT|}{dt}$ = 0.5 K min$^{-1}$ with sample masses between 5 and 15 mg. Calorimetric measurements under applied pressure used sample masses of between 45 and 70 mg and were performed using a Setaram μDSC7 Evo with an applicable hydrostatic pressure limit of up to 1200 bar, using high-purity (99.9999%) nitrogen gas as the pressure-transmitting medium. A temperature-ramping rate of $\frac{|dT|}{dt}$ = 1.2 K min$^{-1}$ was used to obtain isobaric calorimetric signals, $(\frac{dq}{|dT|})_p$, within a temperature range 260 K < $T$ < 340 K, which were then background-subtracted. The integral areas of the residual transition peaks were then obtained to evaluate the barocaloric effect in the samples quasi-directly using

$$|\Delta S_{\mathrm{it}}|(T, p_{\mathrm{i}} \to p_{\mathrm{f}}) = \int_{T_1}^{T} \left[ \left( \frac{\mathrm{d}q(T', p_{\mathrm{f}})}{|\mathrm{d}T|} \right)_p - \left( \frac{\mathrm{d}q(T', p_{\mathrm{i}})}{|\mathrm{d}T|} \right)_p \right] \frac{\mathrm{d}T'}{T'}, \quad (3)$$

where $|\Delta S_{\mathrm{it}}|(T, p_i \to p_f)$ is the isothermal entropy change driven by changes in pressure from $p_i$ (initial pressure) to $p_f$ (final pressure) at temperature $T$. $T_1$ is an arbitrary starting temperature selected below the phase transition at 0 applied pressure and $T'$ is the dummy variable of integration. We use lowercase $q$ for heat to avoid confusion with the scattering variable $Q$ in the neutron spectroscopy data.

### Infra-red imaging

Samples were prepared for imaging by pelletising ~ 0.1g of material into a pellet with 13 mm diameter to create a disk of sample < 1mm in thickness. This was placed on a glass slide and heated to 330 K for 5 minutes using a Linkam PE120 Peltier heating and cooling stage. The sample was then allowed to cool at a rate of 5 K min$^{-1}$ and imaged using a Optris PI640 IR microscope camera as it went through its phase transition on cooling. Images were acquired every 2 s and difference images were constructed by subtracting subsequent frames from preceding ones.

### Synchrotron diffraction

Synchrotron powder diffraction was performed on the I11 beamline at Diamond Light Source, UK using $E_{\mathrm{i}}$ = 15 keV. Samples were packed in 0.5 mm diameter borosilicate capillaries. Temperature was controlled using the Cryostream and data were collected continuously with a ramp rate of 1 K min$^{-1}$ with the position sensitive detectors (PSD). All samples were thermally cycled four times *in-situ* before measurement, to account for virgin effects. Rietveld refinement was performed using GSAS-II[61]. The BCT phase of both PG and NPG-PG-PE was modelled using the published[62] crystal structure of isostructural PE as a starting point. Anisotropic strain parameters, $S_{hkl}$, were determined within the GSAS-II software. Temperature dependent diffraction peak broadening parameters, $S_{hkl}$, for $hkl$ reflections were fitted for each sample according to $\Gamma_{\mathrm{hkl}} \sim S_{\mathrm{HKL}} h^H k^K l^L$, where $\Gamma_{hkl}$ is the diffraction peak full width at half maximum[43]. $S_{400}$, $S_{004}$ and $S_{220}$ parameters were required for NPG-PG-PE refinement whereas only $S_{400}$ and $S_{004}$ parameters gave positive, non-negligible and stable values during PG refinement. Supplementary Fig. 24 shows a comparison of using anisotropic verses isotropic strain models. For the FCC phase, true Rietveld refinement was not attempted due to the significant disorder of the molecule. In order to extract the relative fractions of the BCT and FCC phases the scattering power of the atoms in the FCC unit cell was modelled using a dummy carbon atom with the position and thermal parameter allowed to refine, modelling the two peaks visible in the diffractogram. The resulting refined phase fraction of the FCC phase as a function of temperature on heating and cooling was used in the QENS analysis. Transition temperatures were defined as the temperatures at which the sample had transitioned halfway, given by the phase fraction fitted from XRD refinements.

### Quasielastic neutron scattering

Neutron scattering experiments were performed on the IN16B spectrometer at the Institut Laue–Langevin (ILL), France. IN16B is a backscattering spectrometer and was used in a standard configuration with strained Si111 Doppler monochromator and analysers yielding a FWHM energy resolution of 0.75 μeV, dynamic range of $\pm$ 0.028 meV and $Q$-range of 0.19 to 1.89 Å$^{-1}$. QENS measurements were obtained using scan times of 2 hrs. All QENS measurements were obtained on heating. FWS measurements were performed during a temperature ramp of approximately 0.5 K min$^{-1}$ with alternating acquisitions of elastic (30 s) and inelastic intensity at 3 μeV energy transfer (90 s). The sample was thermally cycled five times immediately before being loaded into a measurement can to account for virgin effects. Approximately 0.5 g of sample was loaded into an aluminium can with annular geometry for measurements on IN16B. Empty can and vanadium standard measurements were acquired using the same sample geometry and used to correct the sample data. Spectrometer resolution measurements were obtained at 2 K.

All data were analysed using Mantid v6.6.0 software[63]. Full details on QENS/FWS analysis and fitting can be found in Supplementary Information Note 3.

## Author Contributions (alphabetical)

Conceptualisation: D.B., D.A.M., F.R.B.; Investigation: M.A., C.B., D.B., E.T.C., M.D., L.G., A.P., F.R.B., C.W.; Formal Analysis: M.A., C.B., D.B., M.D., A.P., F.R.B., C.W.; Writing – original draft: D.B., D.A.M., F.R.B.; Writing – review and editing: all authors; Visualisation; C.B., M.D., D.A.M., F.R.B.; Supervision: D.B., P.L., D.A.M., X.M.; Funding acquisition: D.B., P.L., D.A.M., X.M, F.R.B.

Software, Data curation, Resources, Validation, Methodology, Project administration: Not relevant.

## Acknowledgements

We acknowledge beam time awarded by the ILL neutron source (proposal 7-02-227). We acknowledge Diamond Light Source for time on I11 under proposal CY40567. 

This work was financially supported by an EPSRC grant (EP/V042262/1), a Carnegie Research Incentive Grant (RIG013328) and SGR-00343 Project (Catalonia), by grant PID2023-146623NB-I00 funded by MICIU and by ERDF/EU and is part of Maria de Maeztu Units of Excellence Programme CEX2023-001300-M funded by MCIN/AEI (10.13039/501100011033). We acknowledge an EPSRC strategic equipment funding grant (EP/X030083/1).

### Data Availability

The high-pressure calorimetry, x-ray diffraction and infra-red imaging data that support the findings of this study are available from the following data doi: http://dx.doi.org/10.5525/gla.researchdata.2130. The quasielastic neutron scattering data can be found at the following data doi: https://doi.ill.fr/10.5291/ILL-DATA.7-02-227.

### Competing Interests

The authors declare no competing interests.

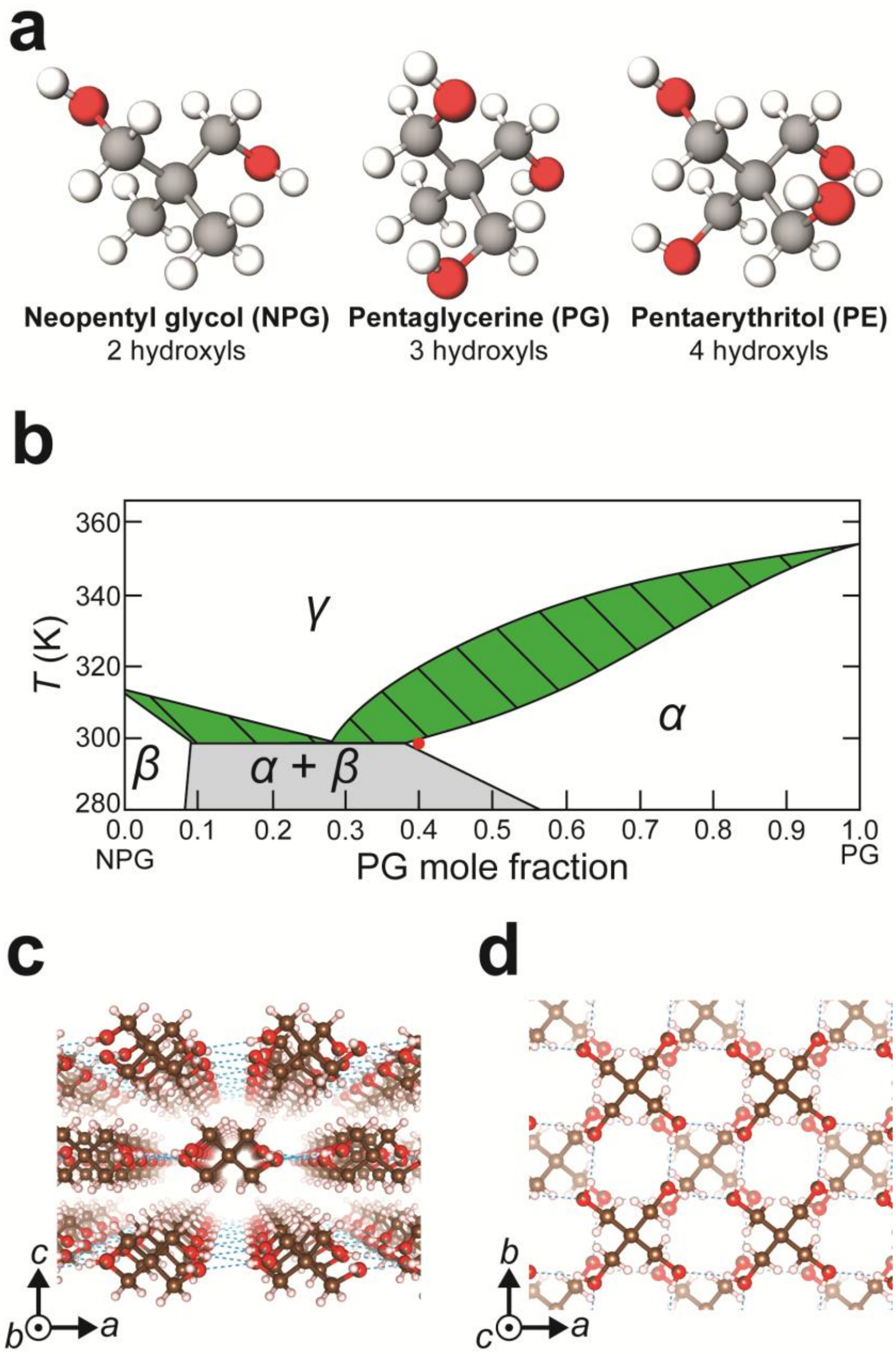

**Fig. 1 | Structures of neopentyl plastic crystals and their solid solutions.** (**a**) Molecular schematics of neopentyl plastic crystals, each tetrahedral molecule differing only in the number of hydroxyl functional groups. (**b**) NPG-PG binary phase diagram adapted from calculated data in the literature[32,33]. $\alpha$, $\beta$, and $\gamma$ correspond to BCT, monoclinic and FCC crystal structures, respectively. The grey section indicates the compositions where the ordered crystal phases are immiscible. The green hatched section indicates the phase coexistence region during ordered crystal to plastic crystal phase transitions. The red dot shows the target binary composition in this work. **c**,**d**) Structural schematics of the BCT structure from two orthogonal perspectives. Hydrogen bonds form layers of molecules in the $ab$ plane with van der Waals forces between layers along the $c$ direction.

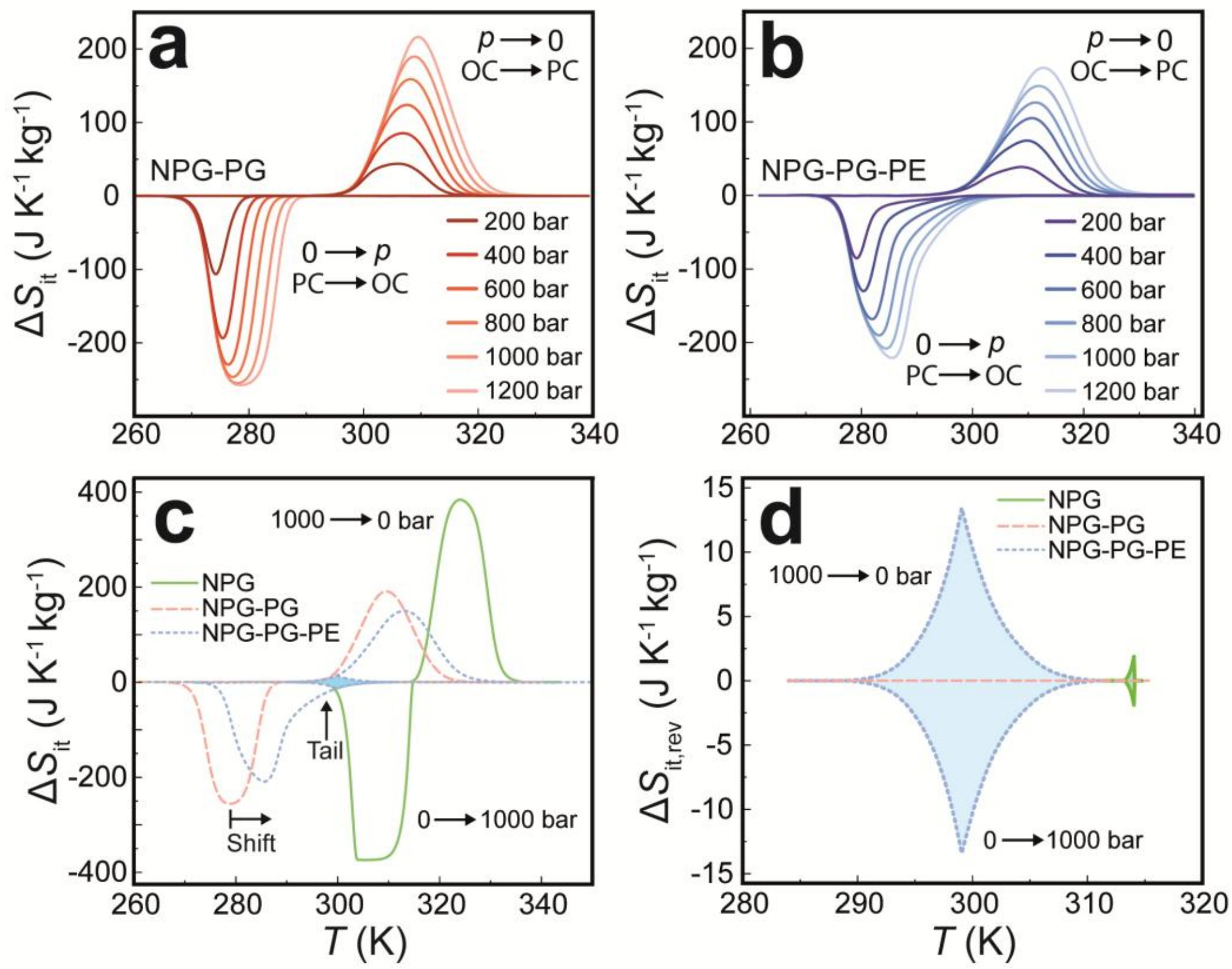

**Fig. 2 | Comparing the barocaloric effect in NPG-PG, NPG-PG-PE and NPG**. Isothermal entropy changes ($\Delta S_{it}$) driven by applying ($0 \rightarrow p$) and removing ($p \rightarrow 0$) pressure in (**a**) NPG-PG and (**b**) NPG-PG-PE. (**c**) Comparison of entropy changes between pure NPG, NPG-PG-PE and NPG-PG when driven by $|\Delta p|$ = 1000 bar. (**d**) Comparison of the reversible entropy change regime between NPG, NPG-PG and NPG-PG-PE when driven by $|\Delta p|$ = 1000 bar.

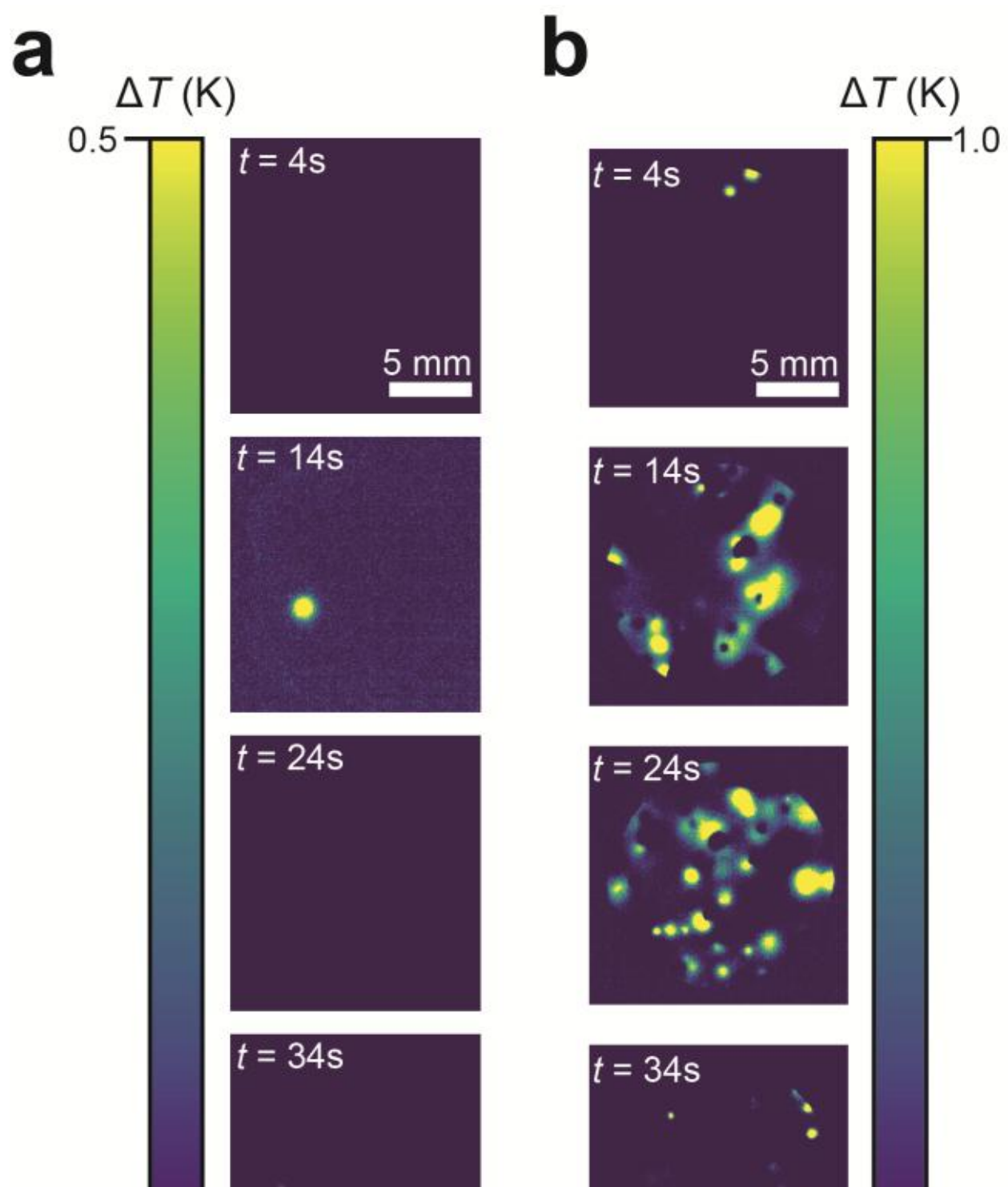

**Fig. 3 | Infra-red imaging of nucleation events.** Infra-red difference images frames of (**a**) NPG-PG and (**b**) NPG-PG-PE on cooling through the PC→OC phase transition. Bright regions indicate nucleation events and phase transition propagation. Cooling rate was 5 K min$^{-1}$. Scale bar is 5 mm.

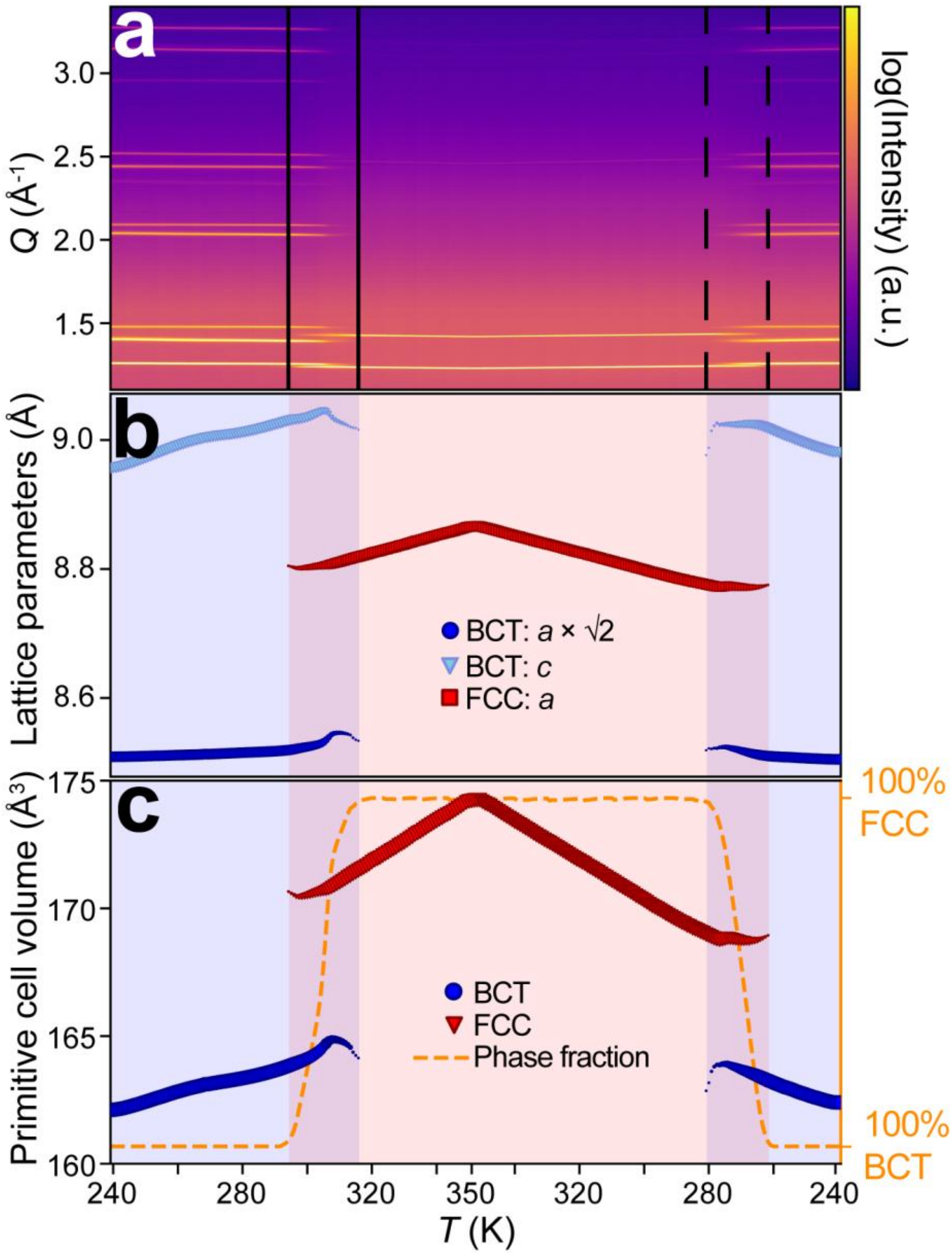

**Fig. 4 | Synchrotron powder x-ray diffraction data for NPG-PG-PE.** (**a**) Full powder XRD data for NPG-PG-PE on heating and cooling. (**b**) Thermal variation of the refined lattice parameters throughout a heating and cooling cycle. BCT lattice parameters at 246 K: $a$ = 6.0170 Å, $c$ = 8.9642 Å. FCC lattice parameters at 319 K $a$ = 8.8234 Å. To facilitate a direct comparison between related vectors, the BCT $a$-parameter is multiplied by a factor of $\sqrt{2}$. (**c**) The primitive unit cell volume determined from the lattice parameters. The dashed orange line shows the FCC phase fraction as a function of temperature, with the shaded red region indicating the phase transition region on heating and cooling. This clearly demonstrates the extended phase coexistence region from below 300 K to nearly 320 K on heating.

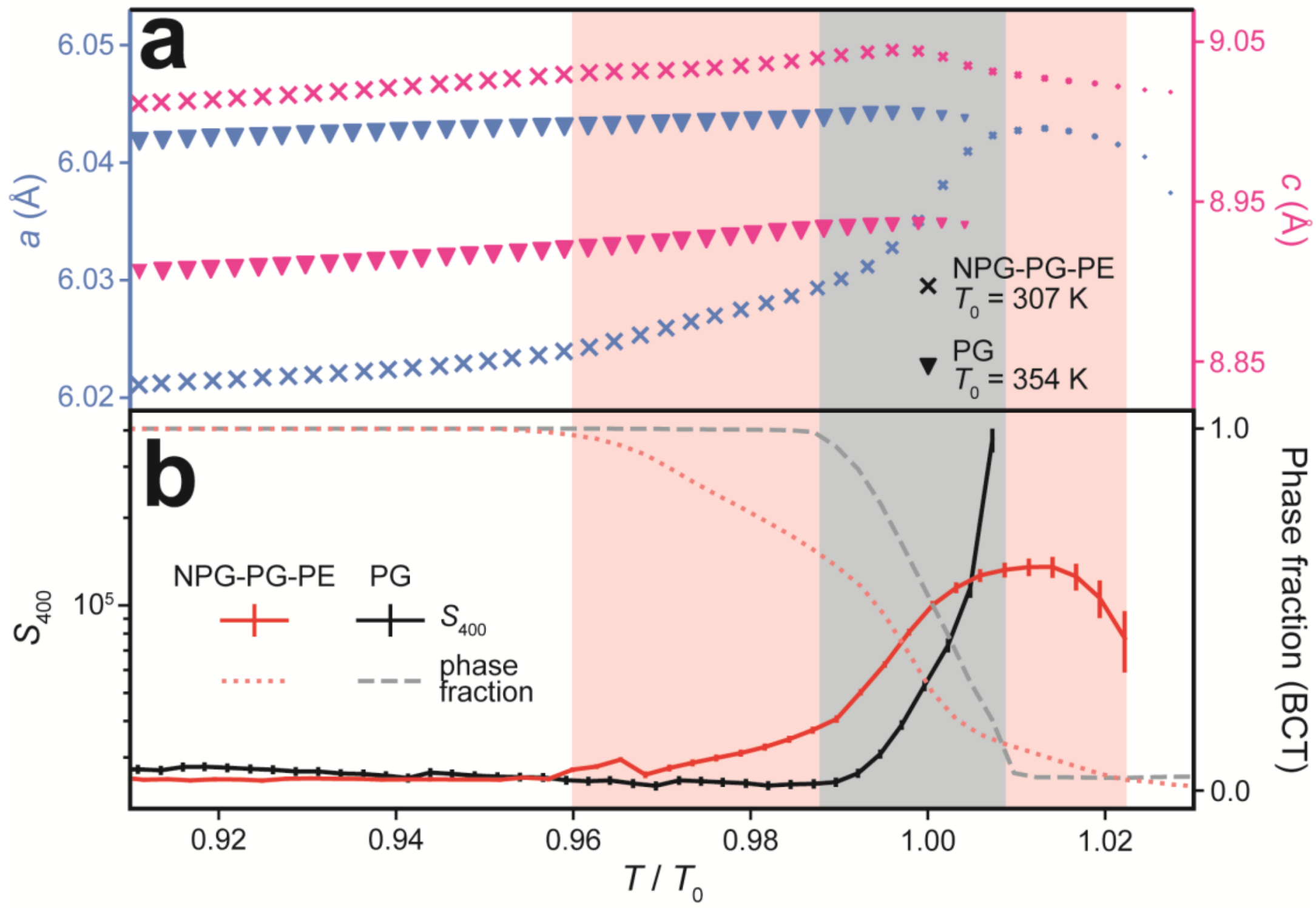

**Fig. 5 | Comparison of structural changes between NPG-PG-PE and PG as the OC→PC phase transition is approached.** (**a**) NPG-PG-PE and PG BCT lattice parameters as a function of reduced temperature, $\frac{T}{T_0}$, showing $a$ (blue, left axis) and $c$ (pink, right axis) as samples are heated through their phase transitions. The relative sizes of the markers encode the phase fraction. (**b**) Anisotropic strain parameter $S_{400}$ and phase fraction as a function of $\frac{T}{T_0}$. The data scale for PG has been increased by a factor of four. Red and grey shaded sections indicate the phase coexistence regions for NPG-PG-PE and PG, respectively.

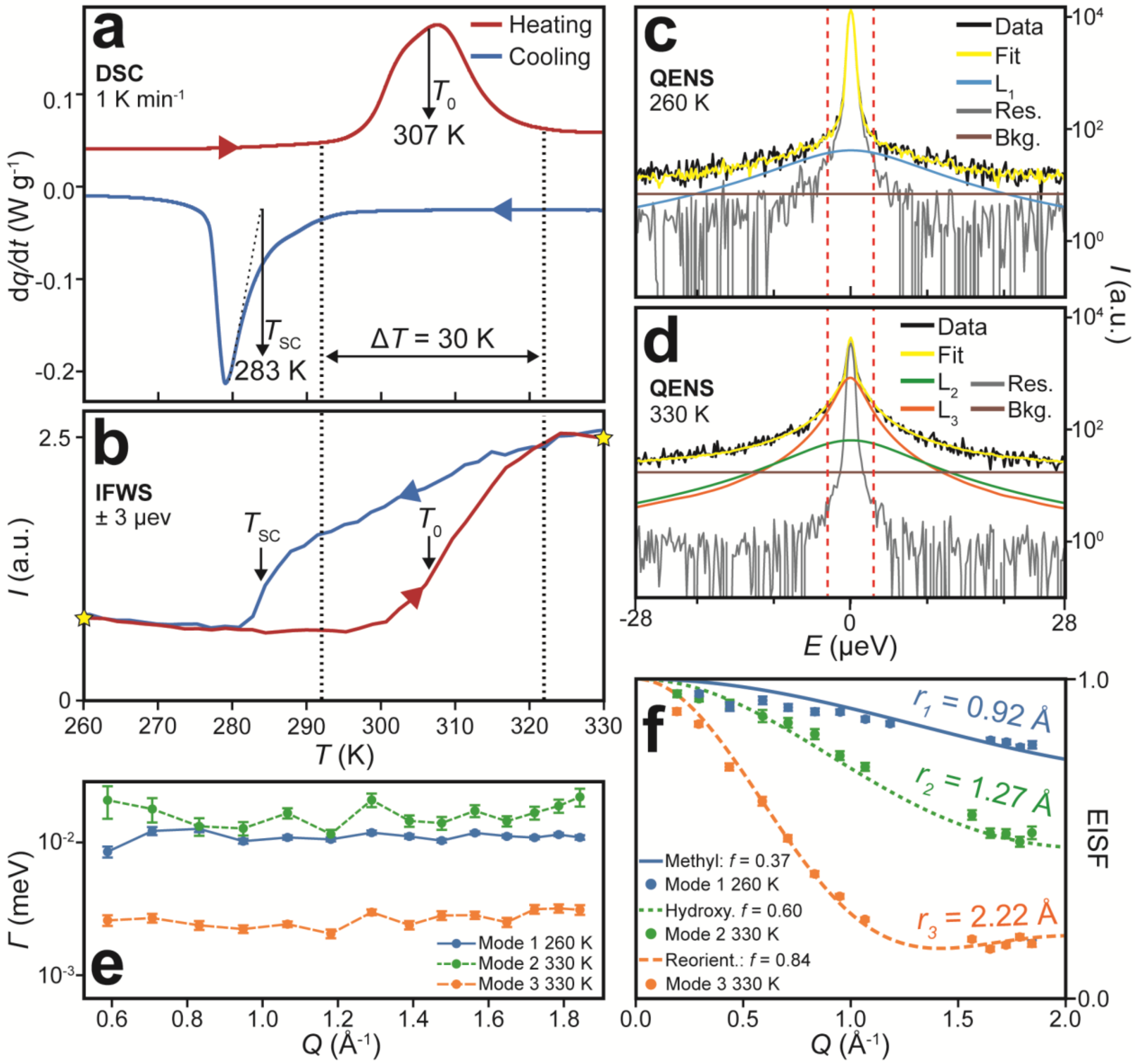

**Fig. 6 | Comparison of calorimetry and neutron spectroscopy data for NPG-PG-PE.** (**a**) Ambient pressure calorimetry data showing the latent heat peaks associated with the phase transition on heating ($T_0$) and cooling ($T_{SC}$). The region between the vertical dashed lines highlights the temperature span of the phase transition during heating, as determined from the XRD phase coexistence region. Note: We intentionally use lowercase $q$ to represent heat to avoid confusion with the scattering variable $Q$. (**b**) Quasielastic neutron IFWS measurements on heating and cooling with two 0.75 μeV wide energy windows at ±3 μeV, collecting the momentum transfer range 0.43 $Å^{-1}$ < $Q$ < 1.84 $Å^{-1}$. Yellow stars indicate temperatures at which QENS scans were obtained. **c**,**d**) Representative QENS fits for $Q$ = 0.83 $Å^{-1}$ at (**c**) 260 K and (**d**) 330 K. The red vertical dashed lines correspond to the IFWS measurements at ±3 μeV. (**e**) Mode linewidths ($\Gamma$) as a function of $Q$ for each of the observed modes in the QENS data. (**f**) $\mathrm{EISF}(Q)$ plot for the three identified modes with fitting to rotational geometric models representing methyl, hydroxymethyl and molecular rotations. Data between 1.1 $Å^{-1}$ < $Q$ < 1.5 $Å^{-1}$ were masked due to the presence of structural Bragg peaks in this $Q$-range.

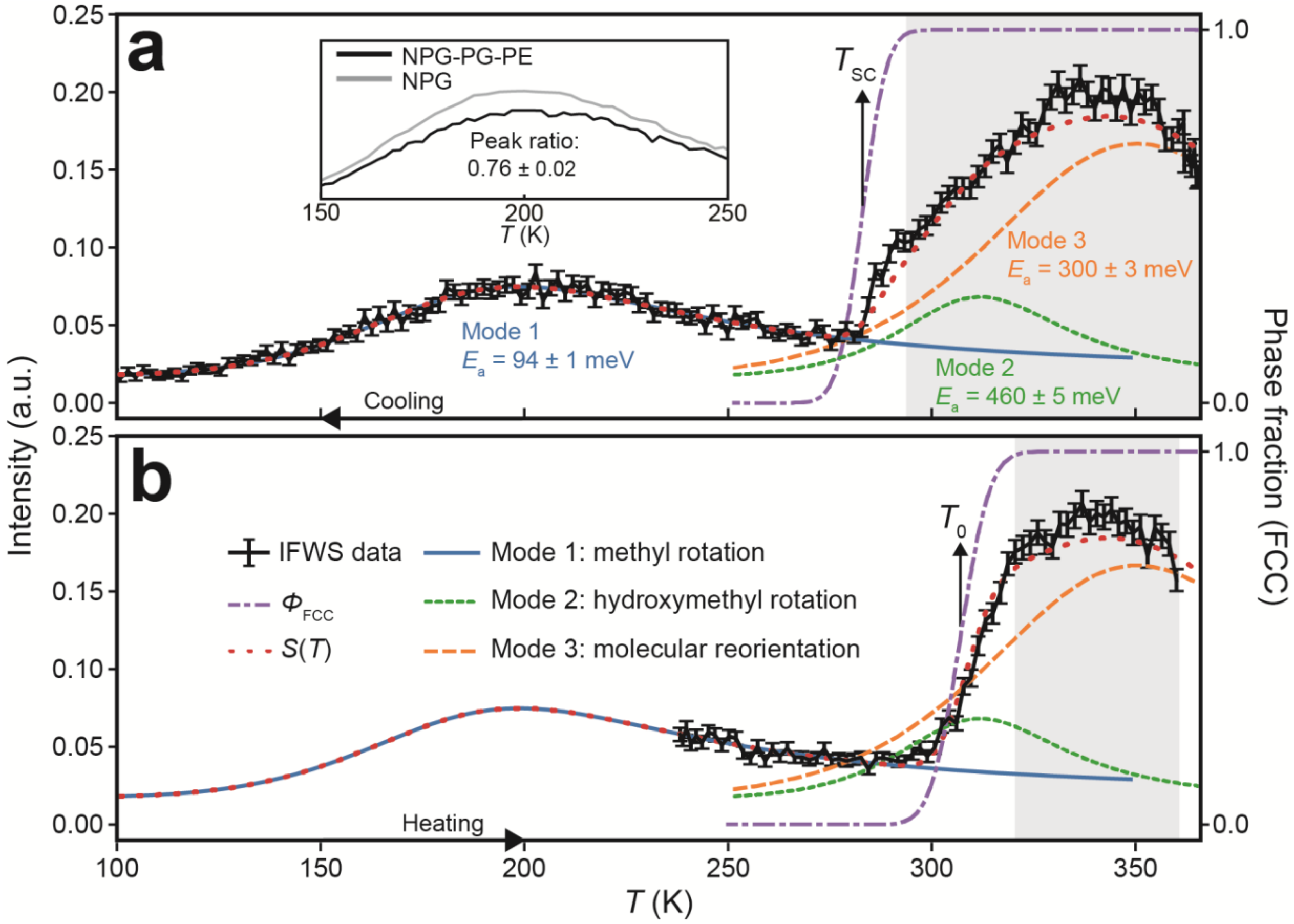

**Fig. 7 | Analysis of inelastic fixed window scan data for NPG-PG-PE.** (**a**) Inelastic fixed window scan ($E$ = $\pm$3 μeV, $Q$ = 0.83 Å$^{-1}$) analysis on cooling from 366 K to 100 K. All inelastic fixed window scan data corresponds to the left axis. The peak centred around 200 K corresponds to the methyl group rotations (solid blue line). The peak centred around 340 K can be fitted with two modes: hydroxymethyl rotation (solid green line) and molecular reorientation (solid orange line). The purple dashed line corresponds to the fitted FCC phase fraction ($\Phi_{\mathrm{FCC}}$, right axis) obtained from XRD and using this it is possible to mix the three modes in the phase coexistence region to provide a fit of $S(T)$ over the entire temperature range (red dotted line). The inset plot compares the methyl rotation peak amplitude between the solid solution and pure NPG. (**b**) Inelastic fixed window scan ($Q$ = 0.83 Å$^{-1}$) on heating from 240 K to 360 K with the fits obtained from the cooling data in (**a**). Shaded regions indicate the temperatures at which $\Phi_{\mathrm{FCC}}$ > 0.98 and the material was considered fully in the PC phase.

# Supplementary Information

# Enhanced reversible barocaloric effect at low pressure in neopentyl plastic crystal solid solutions

*Frederic Rendell-Bhatti*[*1], Melony Dilshad[2], Celine Beck[1], Markus Appel[3], Alba Prats[4], Eamonn T. Connolly[5], Claire Wilson[6], Lewis Giannelli[1], Pol Lloveras[4], Xavier Moya[2], David Boldrin[1], Donald A. MacLaren[1]*

[1]SUPA, School of Physics & Astronomy, University of Glasgow, Glasgow G12 8QQ, United Kingdom

[2]Department of Materials Science & Metallurgy, University of Cambridge, Cambridge, CB3 0FS, United Kingdom

[3]Institut Laue Langevin, 71 Avenue des Martyrs, 38000 Grenoble, France

[4]Group of Characterization of Materials, Department of Physics and Barcelona Research Center in Multiscale Science and Engineering, Universitat Politecnica de Catalunya, Av. Eduard Maristany 16, 08019, Barcelona, Spain

[5]Diamond Light Source, Diamond House, Harwell Science and Innovation Campus, Didcot, Oxfordshire, OX11 0DE, United Kingdom

[6]School of Chemistry, University of Glasgow, Glasgow G12 8QQ, United Kingdom

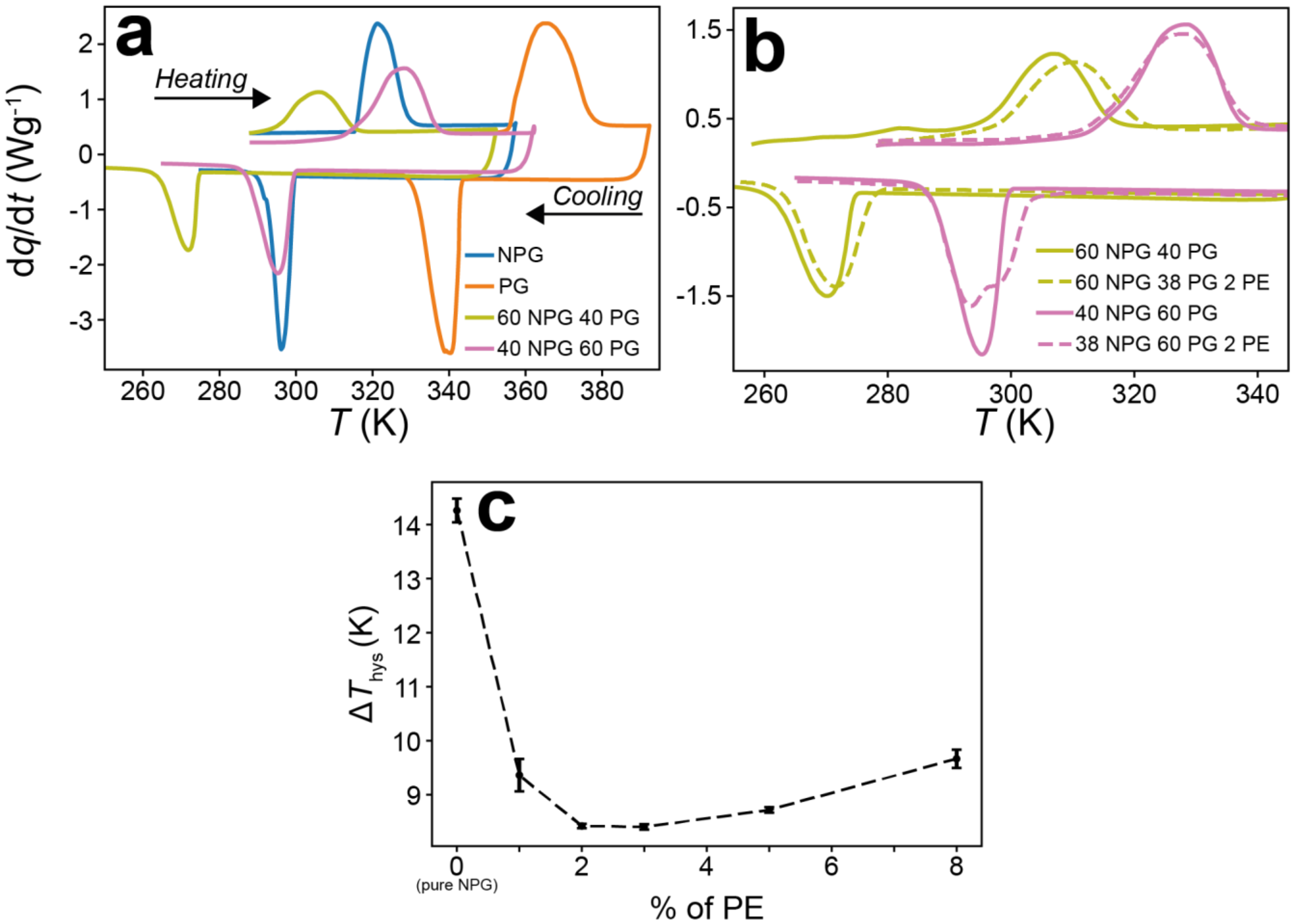

**Supplementary Fig. 1**: Ambient pressure calorimetry of NPG, PG and solid solutions obtained with a heating and cooling ramp rate of 10 K $min^{-1}$. Lowercase *q* has been used for heat to avoid confusion with the neutron spectroscopy scattering constant. All materials were cycled at least 10 times prior to the plots shown to eliminate virgin effects. For both plots in (a,b), exothermic latent heat is in the negative *y*-direction. (a) NPG, PG and two binary solid solutions of 60% NPG 40% PG and 40% NPG 60% PG showing how the binary solid solutions demonstrate significantly different phase transition temperatures and hysteresis compared to the end members. (b) A comparison of the effect of adding a small amount of a third component (2% PE) to the binary solid solutions shown in (a). Although the phase transition temperatures are not significantly altered, there is a reduction in thermal hysteresis for both materials. (c) A comparison of the effect of adding small quantities (1%, 2%, 3%, 5% and 8%) of PE to pure NPG to highlight the effect of small quantities of doping on thermal hysteresis. This plot shows a maximum reduction in thermal hysteresis around 2-3% PE addition, justifying further analysis of the ternary composition in the main text. Hysteresis was calculated as the mean difference in peak temperatures on heating and cooling for the final three cycles from a total of ten, error bars give the standard deviation.

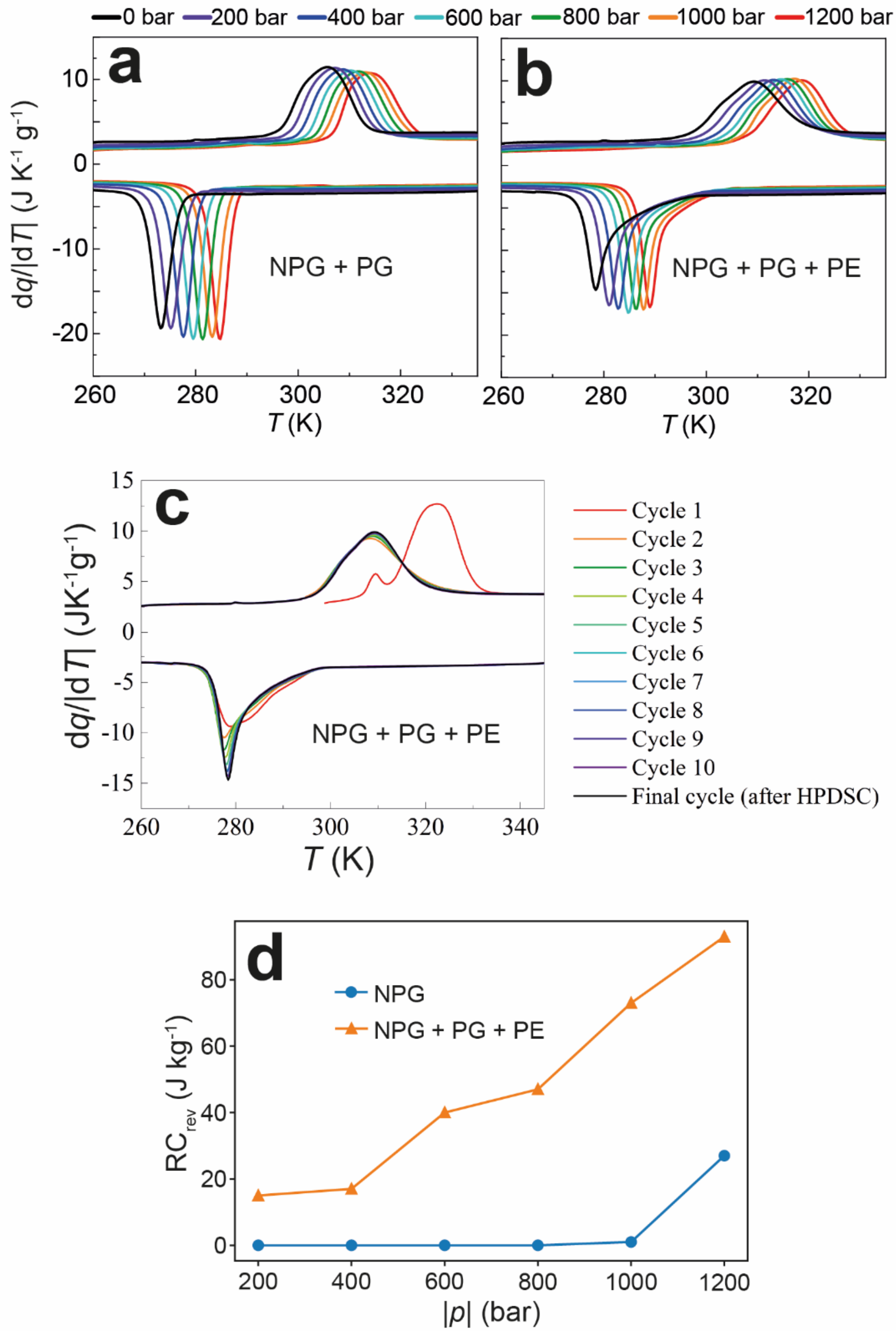

**Supplementary Fig. 2**: High-pressure calorimetry isobaric data obtained using the Setaram calorimeter for (a) 60% NPG 40% PG (NPG+PG) and (b) 60% NPG 38% PG 2% PE (NPG+PG+PE) for different values of increasing pressure. A heating and cooling ramp rate of 1.2 K min$^{-1}$ was used for all data. Lowercase $q$ has been used for heat to avoid confusion with the neutron spectroscopy scattering constant. Nitrogen gas was used as the pressure transmitting medium. (c) Thermal cycling of NPG-PG-PE at ambient pressure before high pressure measurements are obtained. Apart from the virgin effects in cycle 1, NPG-PG-PE shows consistent behaviour across nine successive thermal cycles and shows unchanged behaviour after the high-pressure measurements are taken. (d) The reversible refrigerant capacity of NPG-PG-PE is substantially improved compared to pure NPG at the lower pressures of practicable devices, NPG-PG has zero reversible refrigerant capacity at these pressures and so is omitted from the comparison.

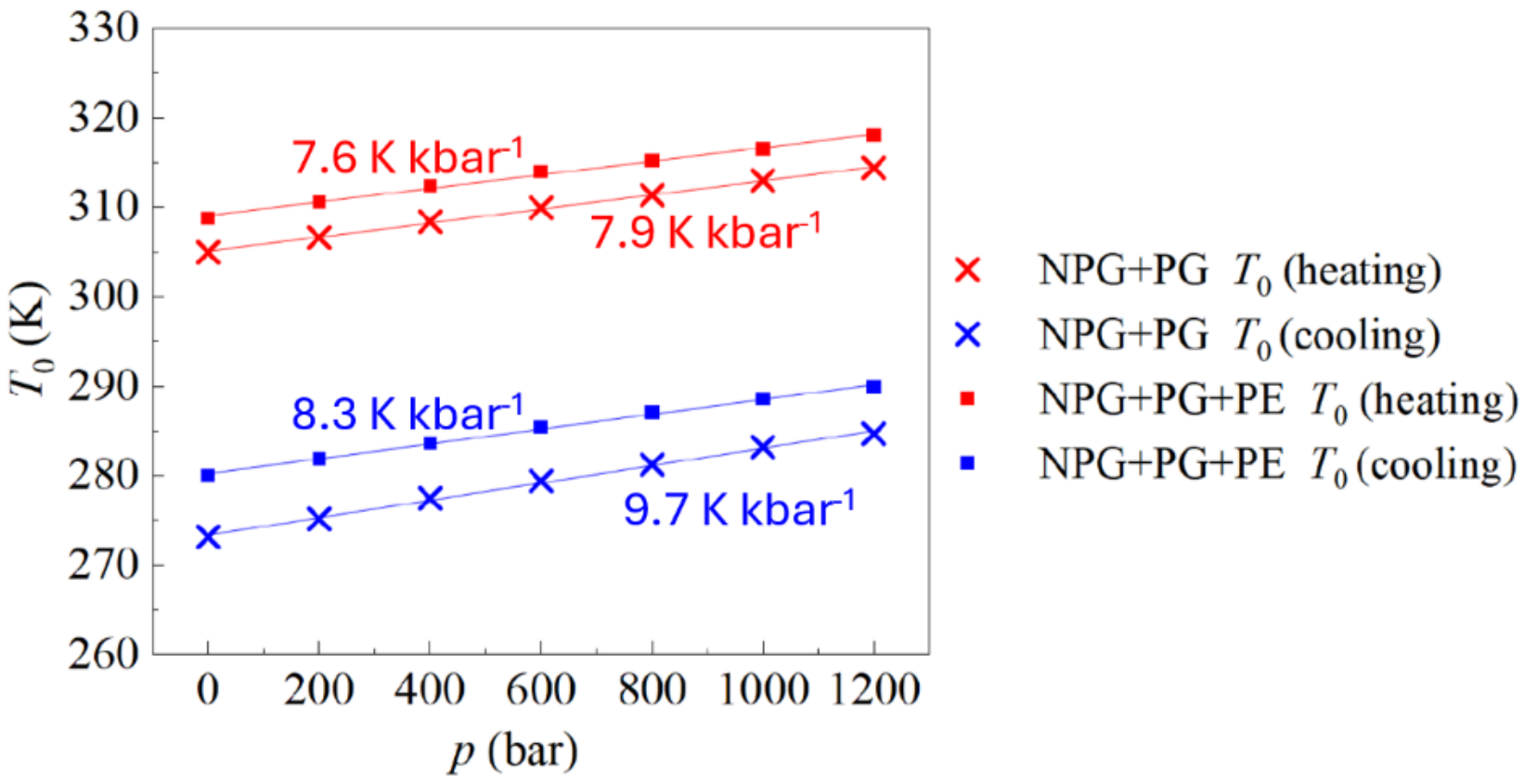

**Supplementary Fig. 3**: Transition temperature as a function of pressure ($dT_0/dp$) on heating and cooling obtained using the Setaram calorimeter for NPG+PG and NPG+PG+PE showing clearly the reduction in peak transition temperature hysteresis for NPG+PG+PE. The annotations give the value of $dT_0/dp$ obtained from each trend. Nitrogen gas was used as the pressure transmitting medium.

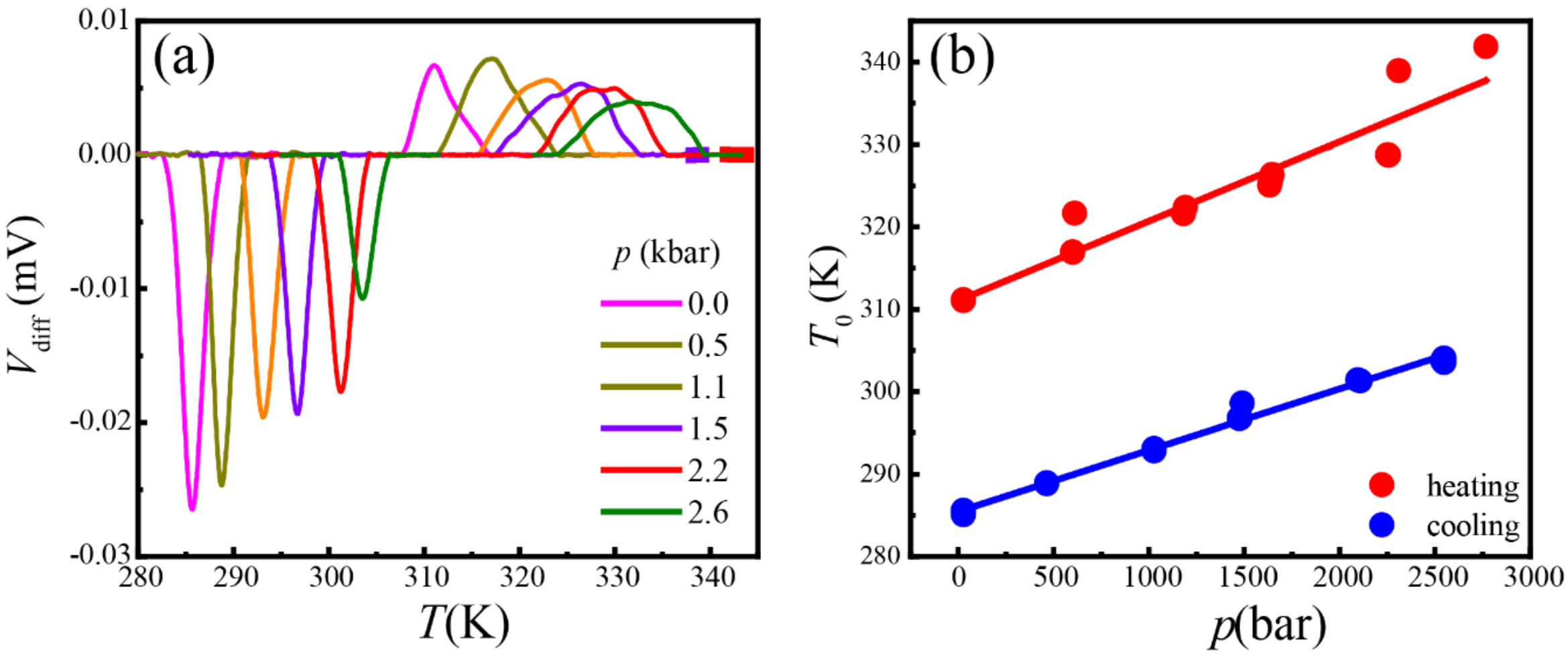

**Supplementary Fig. 4**: (a) Selected thermograms obtained via high-pressure differential thermal analysis and using a custom-made setup with an inert liquid pressure transmitting medium. Pressures in the legend have been averaged for heating and cooling ramps. (b) Transition temperatures as a function of pressure, determined from the peak maxima. The obtained $dT/dp$ results are 9.65 K kbar$^{-1}$ and 7.45 K kbar$^{-1}$ for heating and cooling data, respectively. The scattering and slight discrepancy with the data obtained using the commercial setup presented in Figs. S2 and S3 may occur due to inhomogeneities arising in large sample amounts and consequent inhomogenous sample environments.

For data in Supplementary Fig. 4, high-pressure differential thermal analysis was performed using a custom-made Cu-Be calorimeter using chromel-alumel Bridgman thermocouples as thermal sensors, operating up to 3 kbar and within the temperature range from 200 K and up to 393 K. Temperature ramps at ~ ±3 K min$^{-1}$ were performed using an external thermostat (Lauda Proline RP 1290). Hundreds of mg of powder mixtures of NPG-PG-PE were encapsulated with an inert liquid (Galden Bioblock Scientist) inside tin capsules. A closed hole was drilled into the capsules to insert the thermocouples. The pressure-transmitting fluid was DW-Therm.M90.200.02. (Huber). Looking at the heating peaks in (a) and the eventual scattering in (b), we conclude that this behaviour can be ascribed to the large amount of sample (a few hundreds of mg) mixing with the pressure transmitting liquid, which may introduce some inhomogeneities in particle size, so regions with slightly different concentrations can give rise to slightly different transition temperatures.

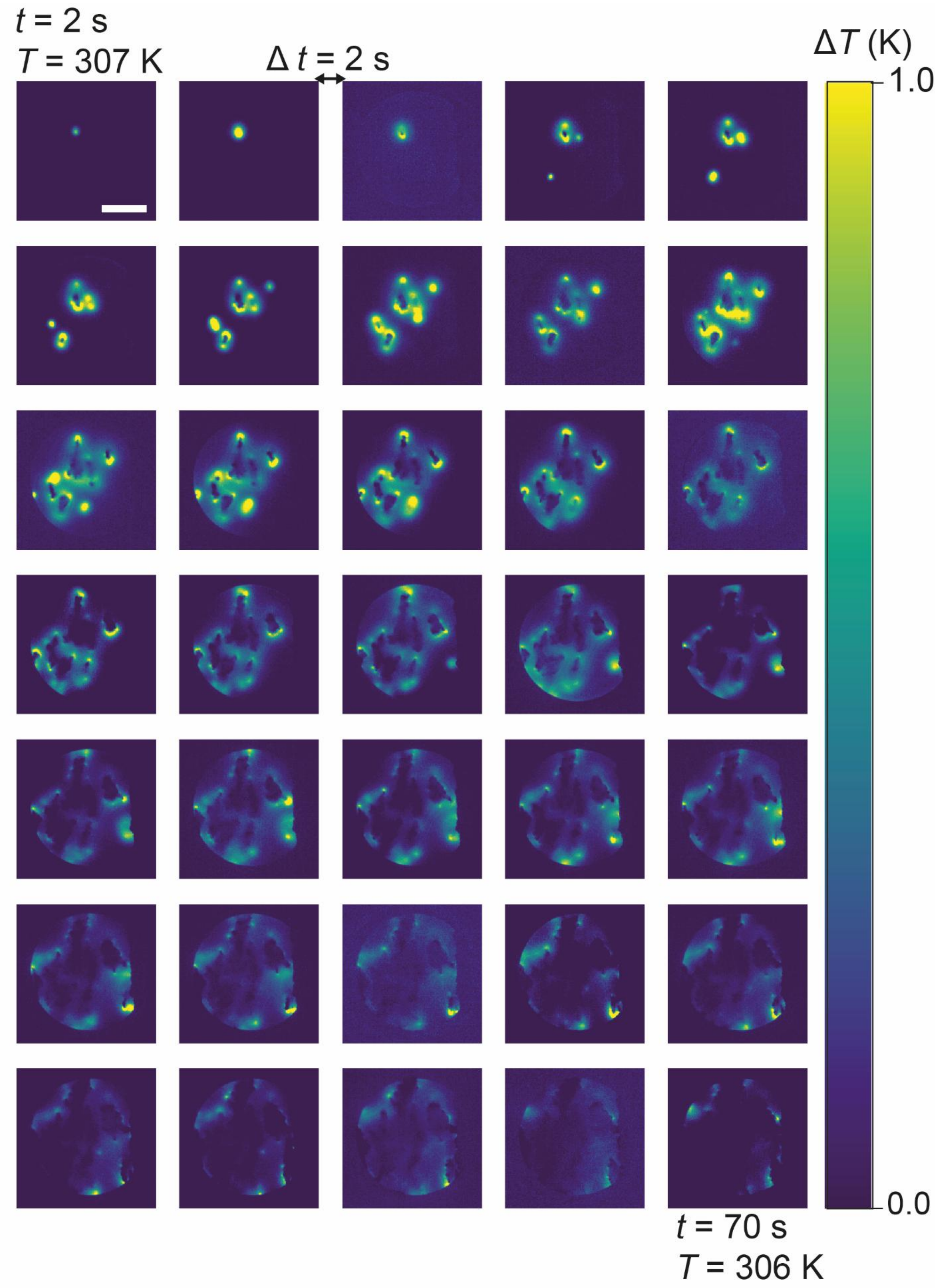

**Supplementary Fig. 5**: Infra-red difference imaging of NPG on cooling through the supercooled PC-OC phase transition with a cooling ramp rate of 1 K min$^{-1}$ and at ambient pressure. As described in the main text, difference images select only those regions that have recently undergone the phase transition and hence are hotter than in previous frames – it therefore picks out points of nucleation and where the phase transition is propagating across the field of view. The sample is a compressed ~0.1g circular disk of 13 mm diameter and approximately 1 mm thick, which is visible as the circular feature in later images. The first nucleation point appears after 2s followed by more between 8 s < $t$ < 16 s, these nucleation points propagating anisotropically through the material representing linear phase transition wave fronts, which could be due to the needle-like crystal habit of NPG[1]. These appear in the images as a bright spot at the end of a darkened region. Scale bar is 5 mm.

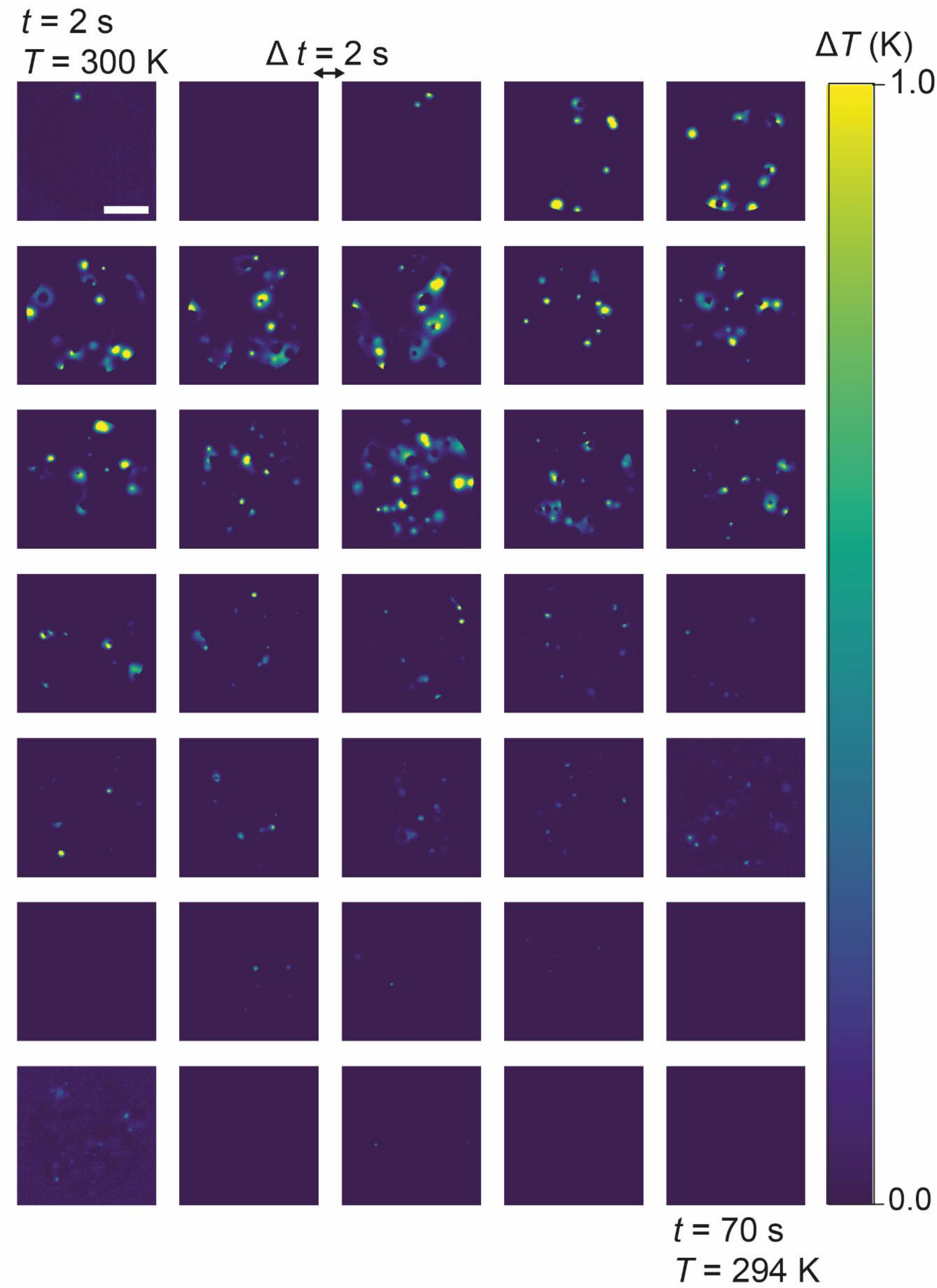

**Supplementary Fig. 6**: Infra-red difference images of NPG-PG-PE on cooling through the supercooled PC-OC phase transition with a cooling ramp rate of 5 K min$^{-1}$ at ambient pressure. Other experimental details, including sample size, are as for Supplementary Fig. 5 and outlined in the Methods Section of the manuscript. There is a significant number of nucleation events within the first ~30s of the transition. The spatial extent of each nucleation point is significantly lower than in pure NPG, shown in Supplementary Fig. 5, and these nucleation events are spreading out radially unlike the linear behaviour in NPG. This may suggest a smaller grain size and greater mosaicity in comparison to the strong preference for needle-like crystals in NPG. Scale bar is 5 mm.

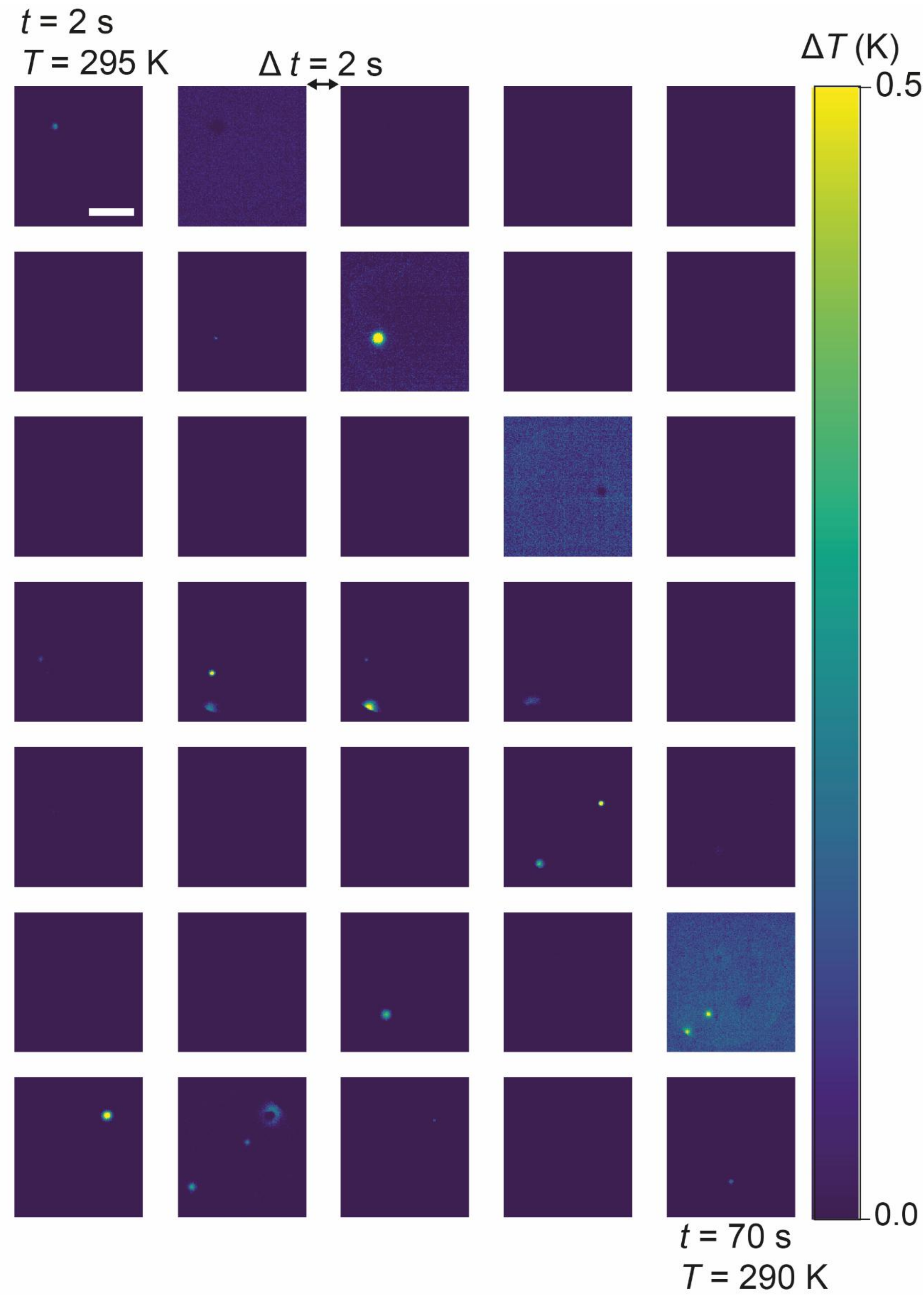

**Supplementary Fig. 7**: Infra-red difference images of NPG-PG on cooling through the supercooled PC-OC phase transition with a cooling ramp rate of 5 K $min^{-1}$at ambient pressure. Other experimental details, including sample size, are again as for Supplementary Fig. 5 and outlined in the Methods Section of the manuscript. The number of nucleation events is evidently much lower than in the NPG-PG-PE sample shown in Supplementary Fig. 6. As with NPG-PG-PE, however, there is no obvious directionality to the propagation of heat emission, suggesting a more isotropic crystal texture than for NPG. Scale bar is 5 mm.

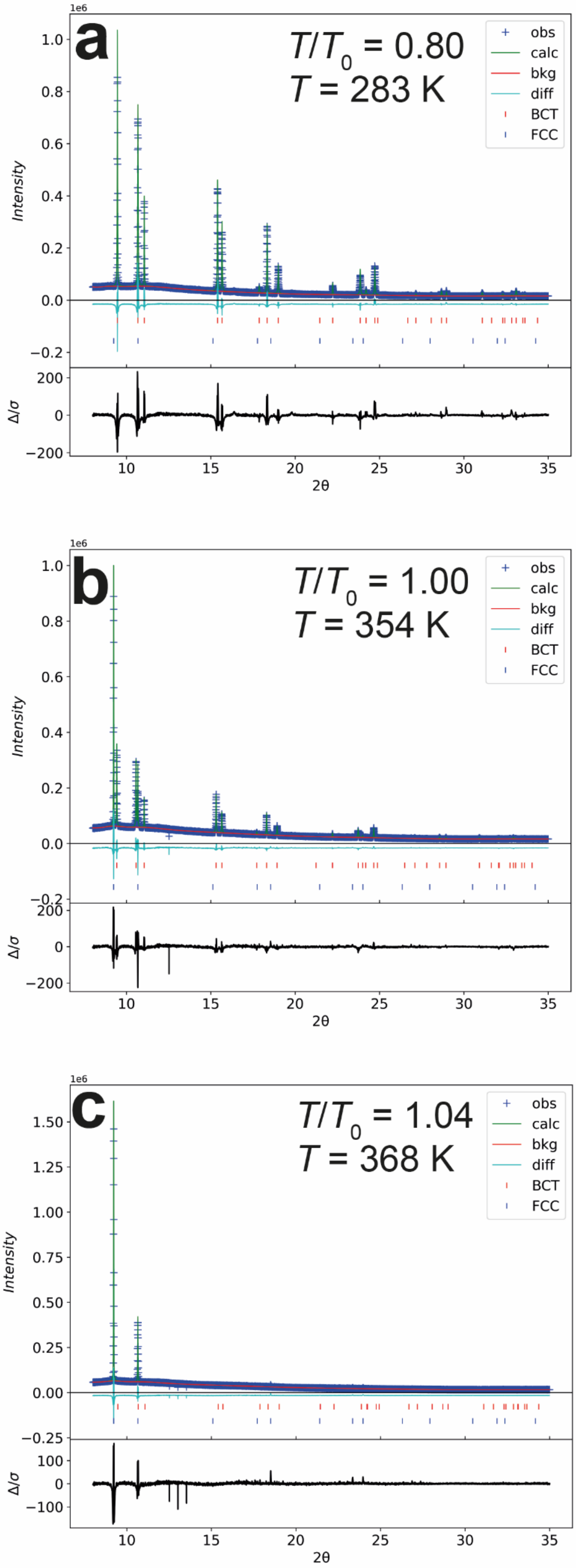

**Supplementary Fig. 8**: Powder synchrotron ($\lambda$ = 0.824875 Å) XRD Rietveld refinements for PG on heating from the OC BCT phase to the PC FCC phase. (a) $T/T_0$ = 0.80, $T$ = 283 K, below the phase transition. BCT with lattice parameters $a$ = 6.0398 Å, $c$ = 8.8451 Å. (b) $T/T_0$ = 1.00, $T$ = 354, during the phase transition. (c) $T/T_0$ = 1.04, $T$ = 368 K, above the phase transition. FCC with lattice parameters $a$ = 8.8658 Å.

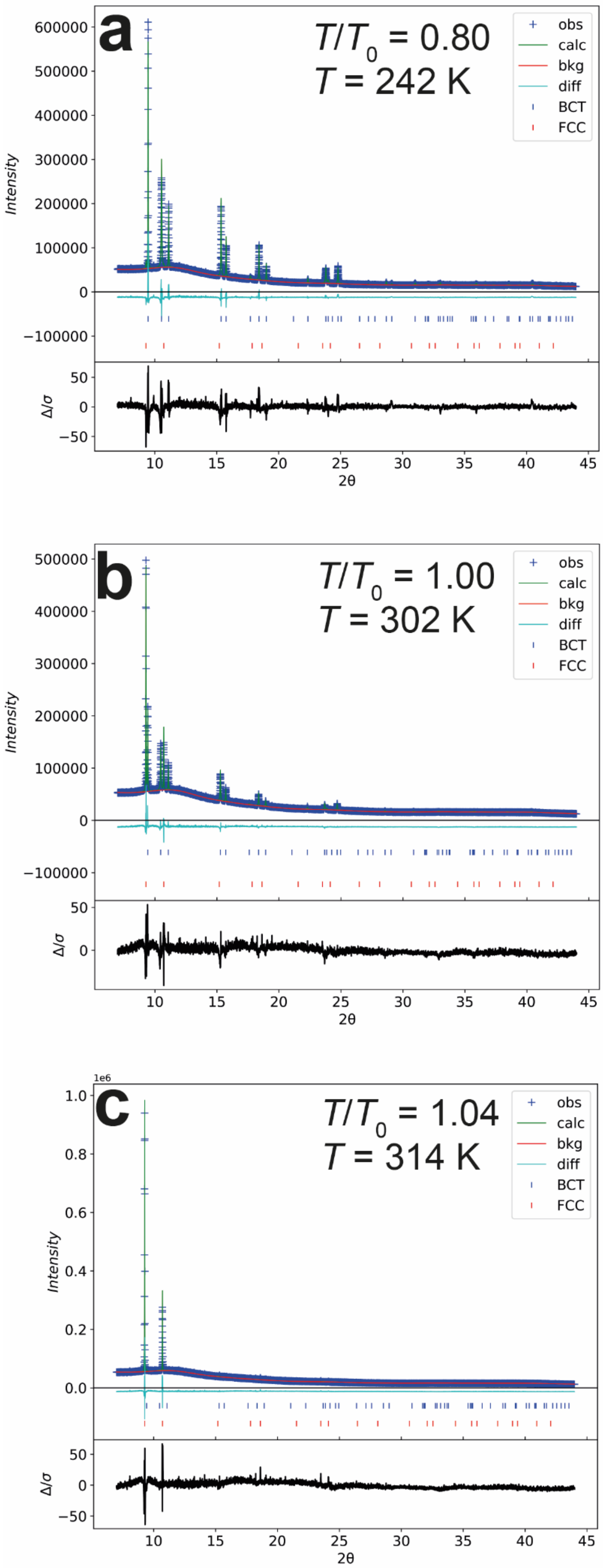

**Supplementary Fig. 9**: Powder synchrotron ($\lambda$ = 0.824875 Å) XRD Rietveld refinements for NPG-PG on heating from the OC BCT phase to the PC FCC phase. (a) $T/T_0$ = 0.80, $T$ = 242 K, below the phase transition. BCT with lattice parameters: $a$ = 6.0089 Å, $c$ = 8.9594 Å. (b) $T/T_0$ = 1.00, $T$ = 302 K, during the phase transition. (c) $T/T_0$ = 1.04, $T$ = 314 K, above the phase transition. FCC with lattice parameters: $a$ = 8.8191 Å.

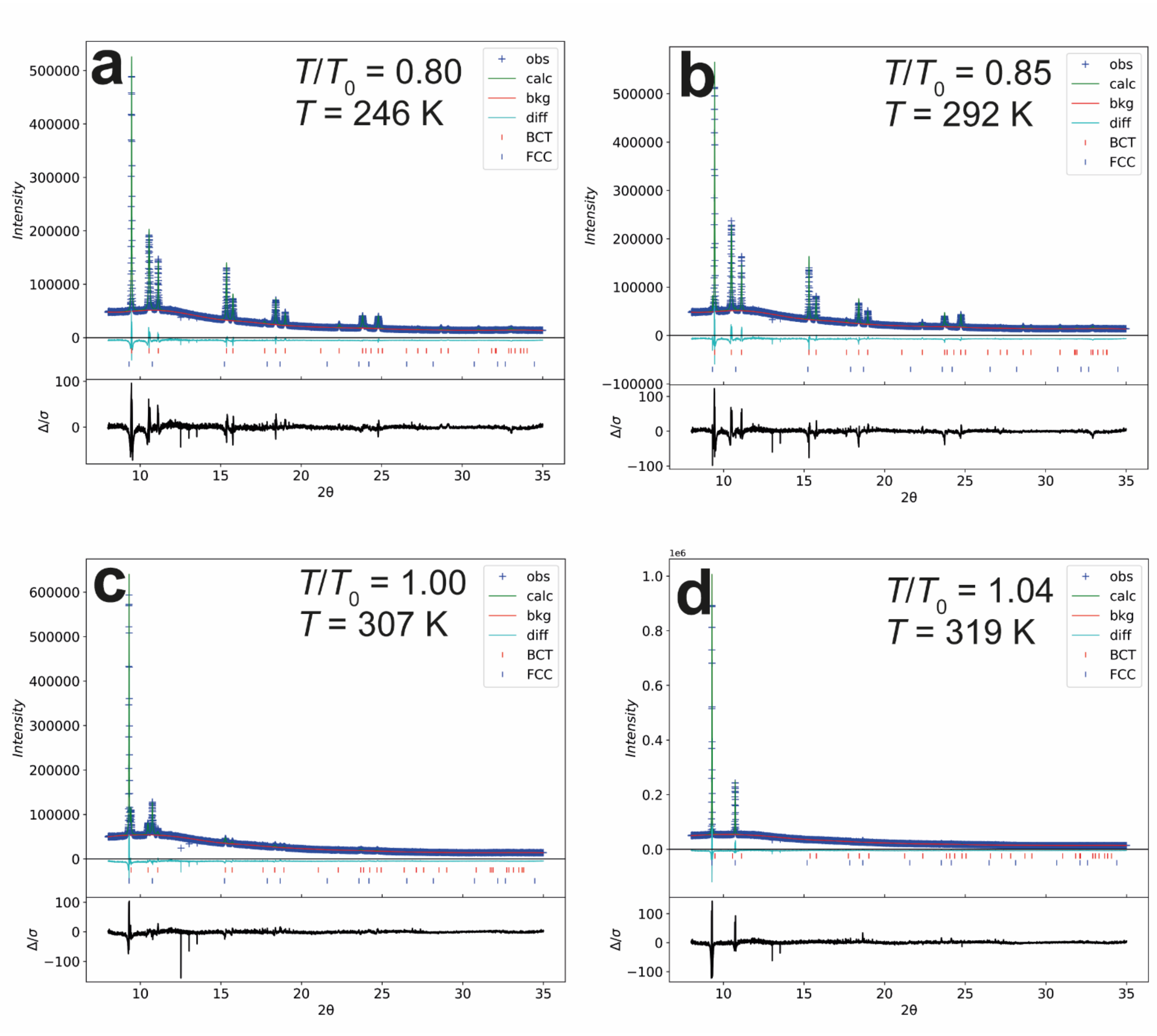

**Supplementary Fig. 10**: Powder synchrotron ($\lambda$ = 0.824875 Å) XRD Rietveld refinements for NPG-PG-PE on heating from the OC BCT phase to the PC FCC phase. (a) $T/T_0$ = 0.80, $T$ = 246 K, below the phase transition. BCT with lattice parameters $a$ = 6.0170 Å, $c$ = 8.9642 Å. (b) $T/T_0$ = 0.95, $T$ = 292 K, below the phase transition. BCT with lattice parameters $a$ = 6.0239 Å, $c$ = 9.0295 Å. (c) $T/T_0$ = 1.00, $T$ = 307 K, during the phase transition. (d) $T/T_0$ = 1.04, $T$ = 319 K, above the phase transition. FCC with lattice parameters $a$ = 8.8234 Å.

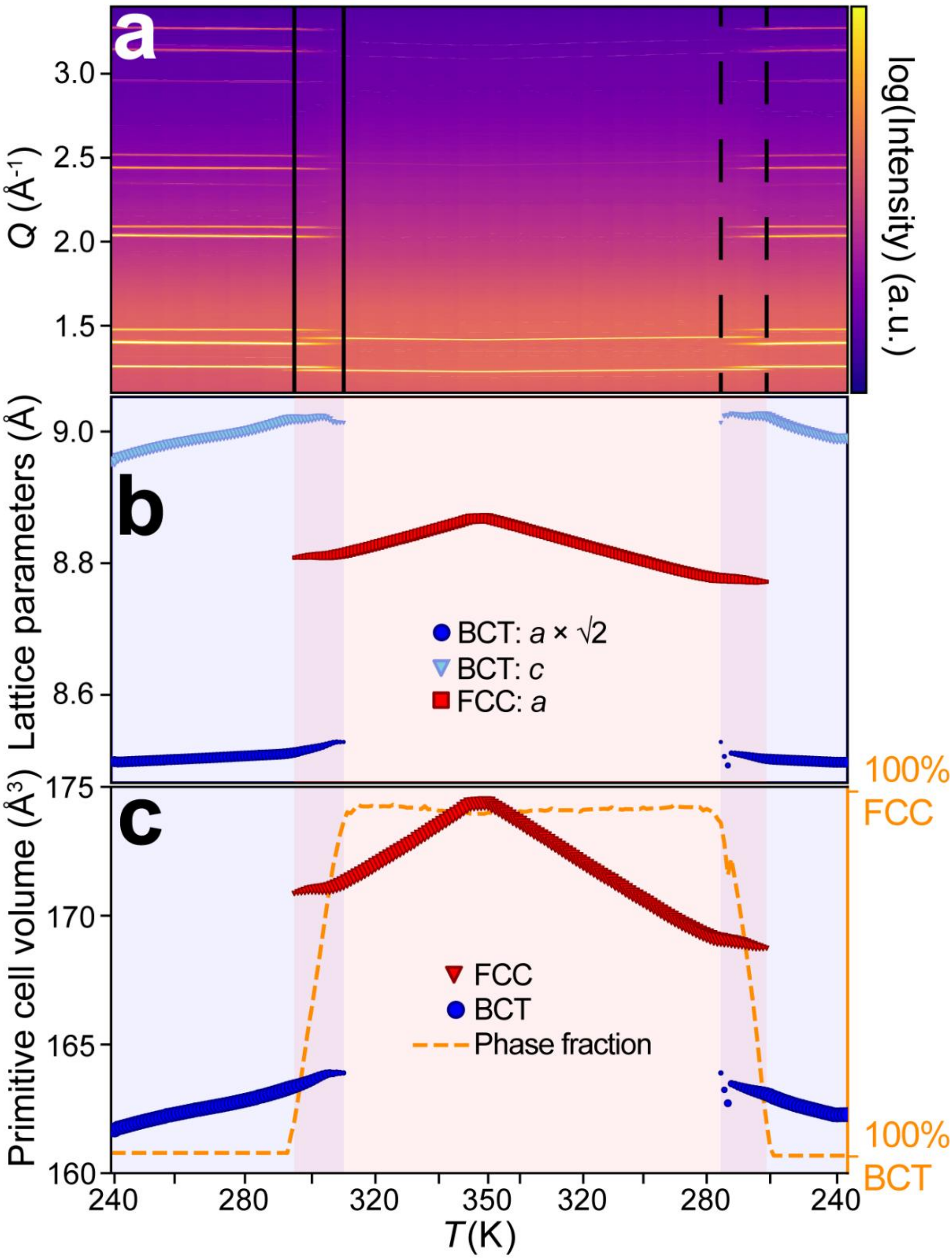

**Supplementary Fig. 11**: (a) Full synchrotron powder XRD data for NPG-PG. Region between solid(dashed) black lines indicates heating(cooling) phase transition. (b) Refined lattice parameters as a function of temperature. The *a* parameter for BCT phase has been multiplied by $\sqrt{2}$. (c) Primitive cell volume and phase fraction as a function of temperature. The shaded blue and red regions indicate the temperatures where the BCT and FCC phases are present, respectively.

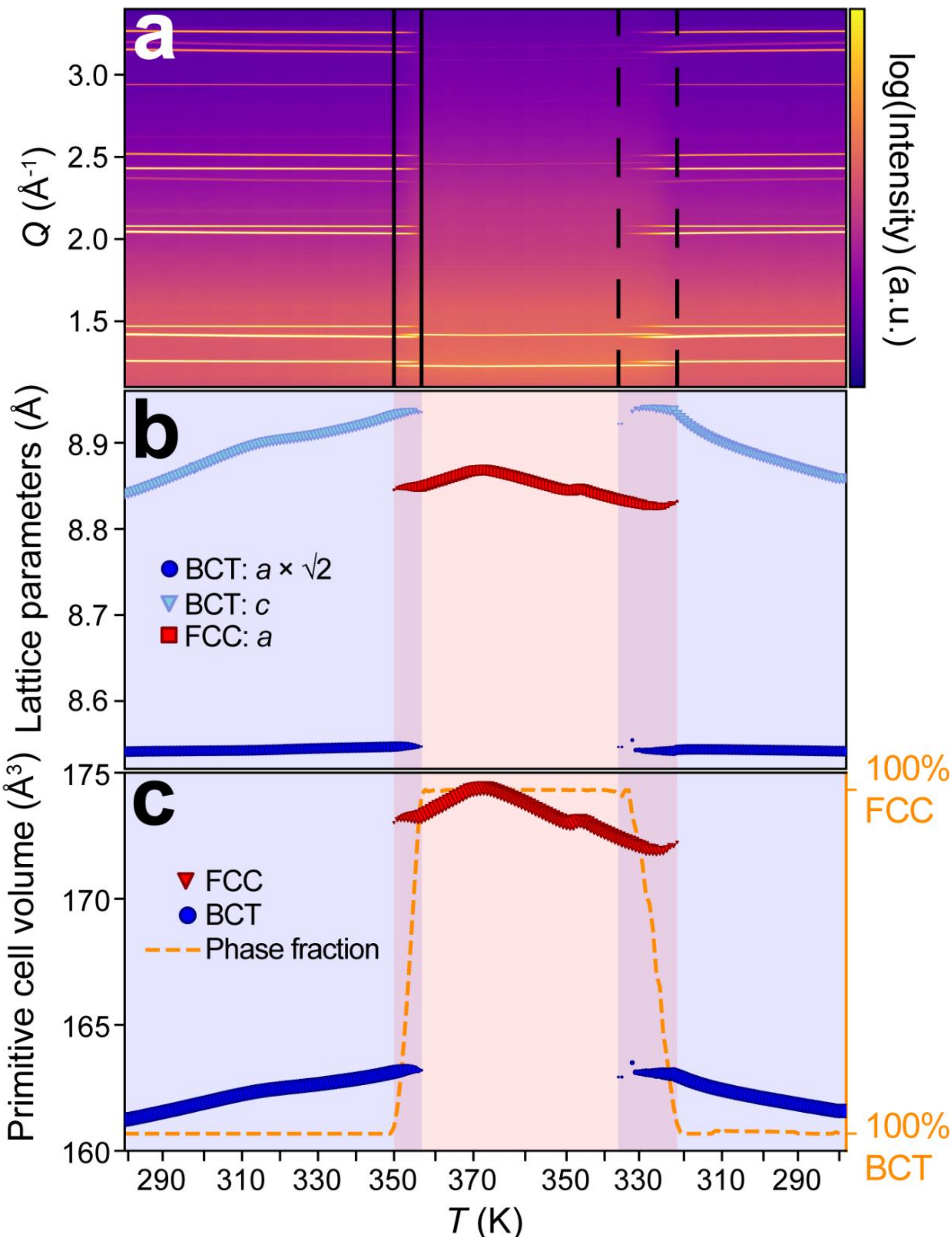

**Supplementary Fig. 12:** (a) Full synchrotron powder XRD data for PG. Region between solid(dashed) black lines indicates heating(cooling) phase transition. (b) Refined lattice parameters as a function of temperature. The *a* parameter for BCT phase has been multiplied by $\sqrt{2}$. (c) Primitive cell volume and phase fraction as a function of temperature. The shaded blue and red regions indicate the temperatures where the BCT and FCC phases are present, respectively.

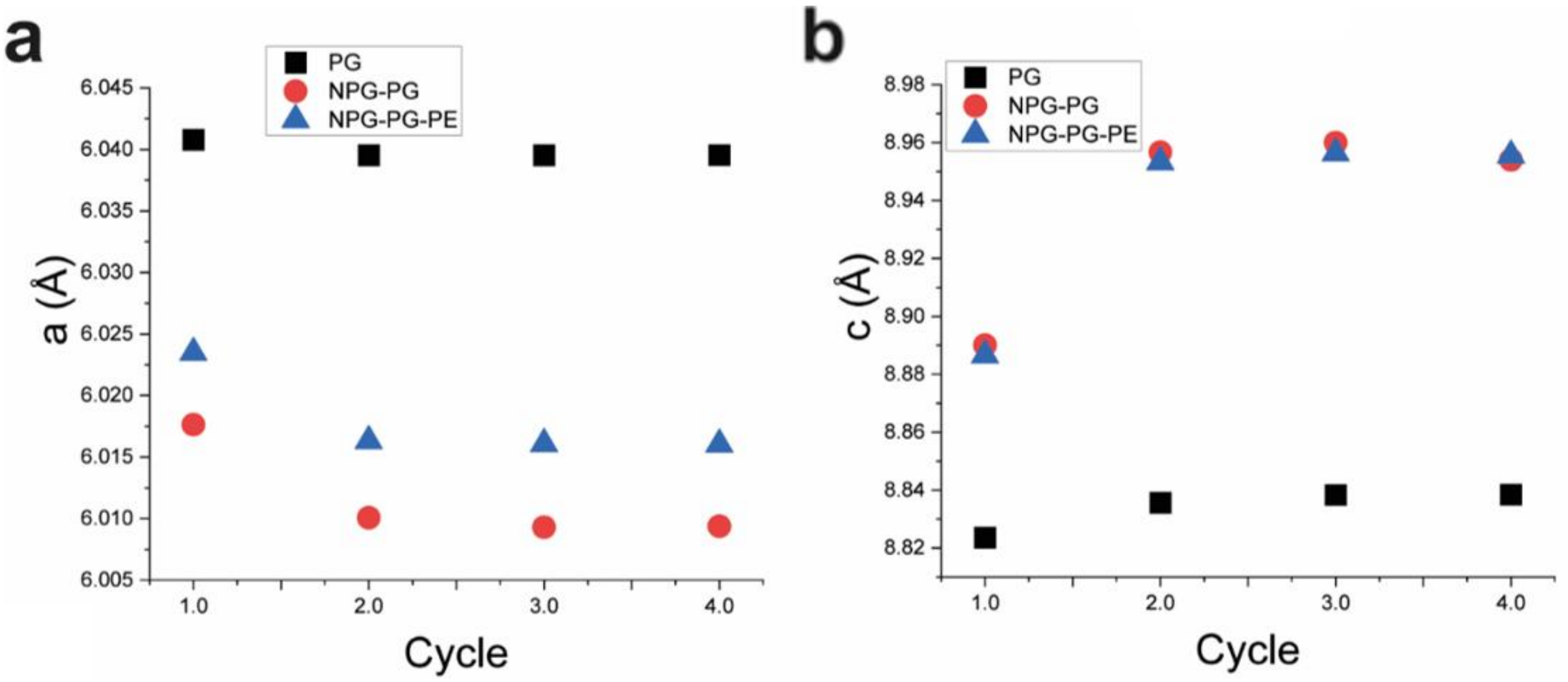

**Supplementary Fig. 13**: BCT lattice parameters for PG at 280 K and NPG-PG, NPG-PG-PE at 240 K for the first four thermal cycles before continuous measurement. This shows that after the first cycle, which demonstrates virgin effects[1], the *a* and *c* lattice parameters stabilise for all three materials.

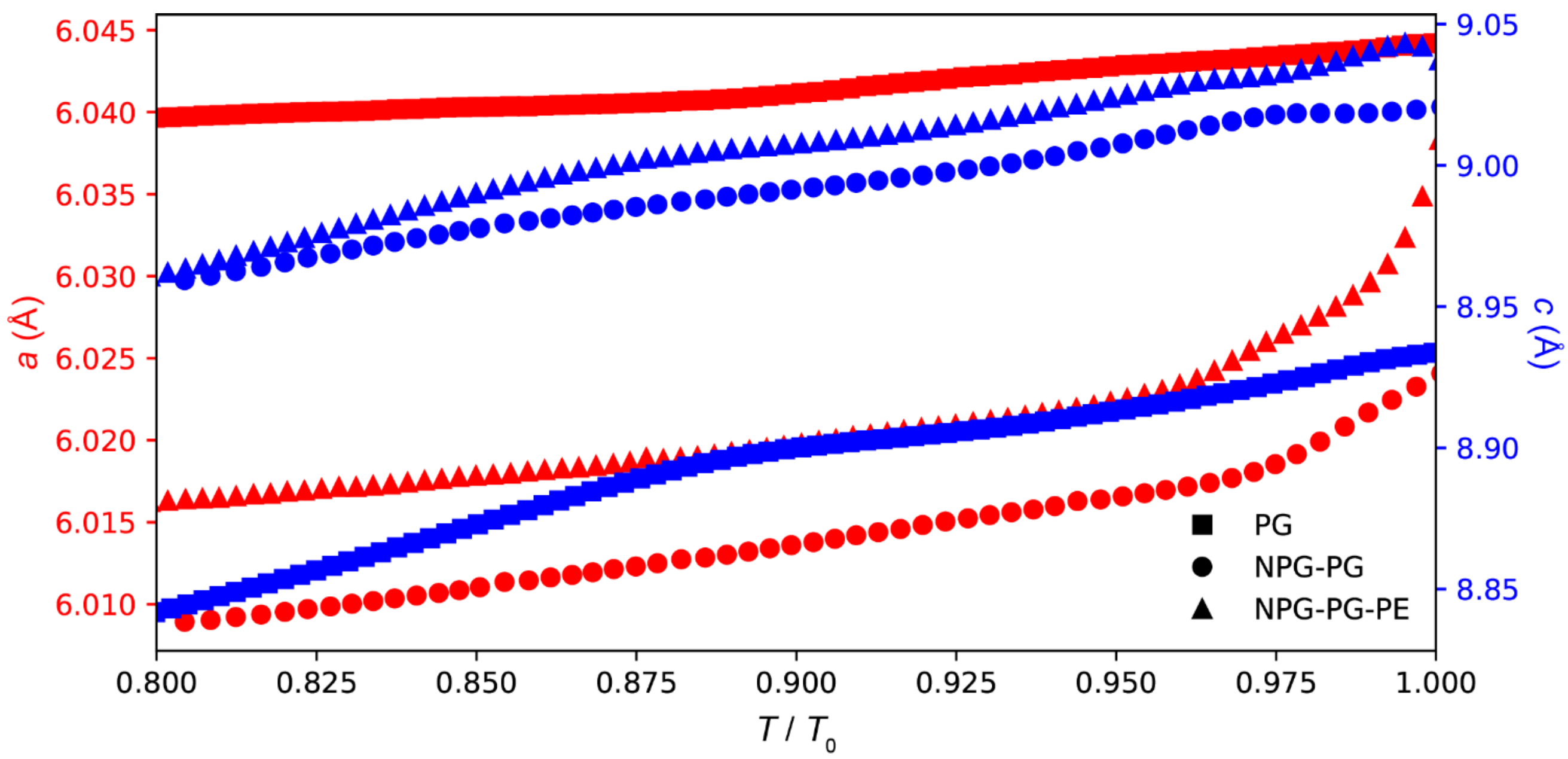

**Supplementary Fig. 14**: BCT lattice parameters for PG, NPG-PG and NPG-PG-PE on heating, showing the *a*-parameters in red (left hand axis) and the *c*-parameters in blue (right hand axis). The data demonstrate that the temperature dependence of the *a* lattice parameter approaching the transition is similar in both NPG-PG and NPG-PG-PE, with increase above the linear trend for reduced temperatures above 0.950. In contrast there is no such variation for PG, nor for any of the *c*-parameters. These variations in lattice constant indicate that the phase transition for the blended materials is broader than that for pure PG.

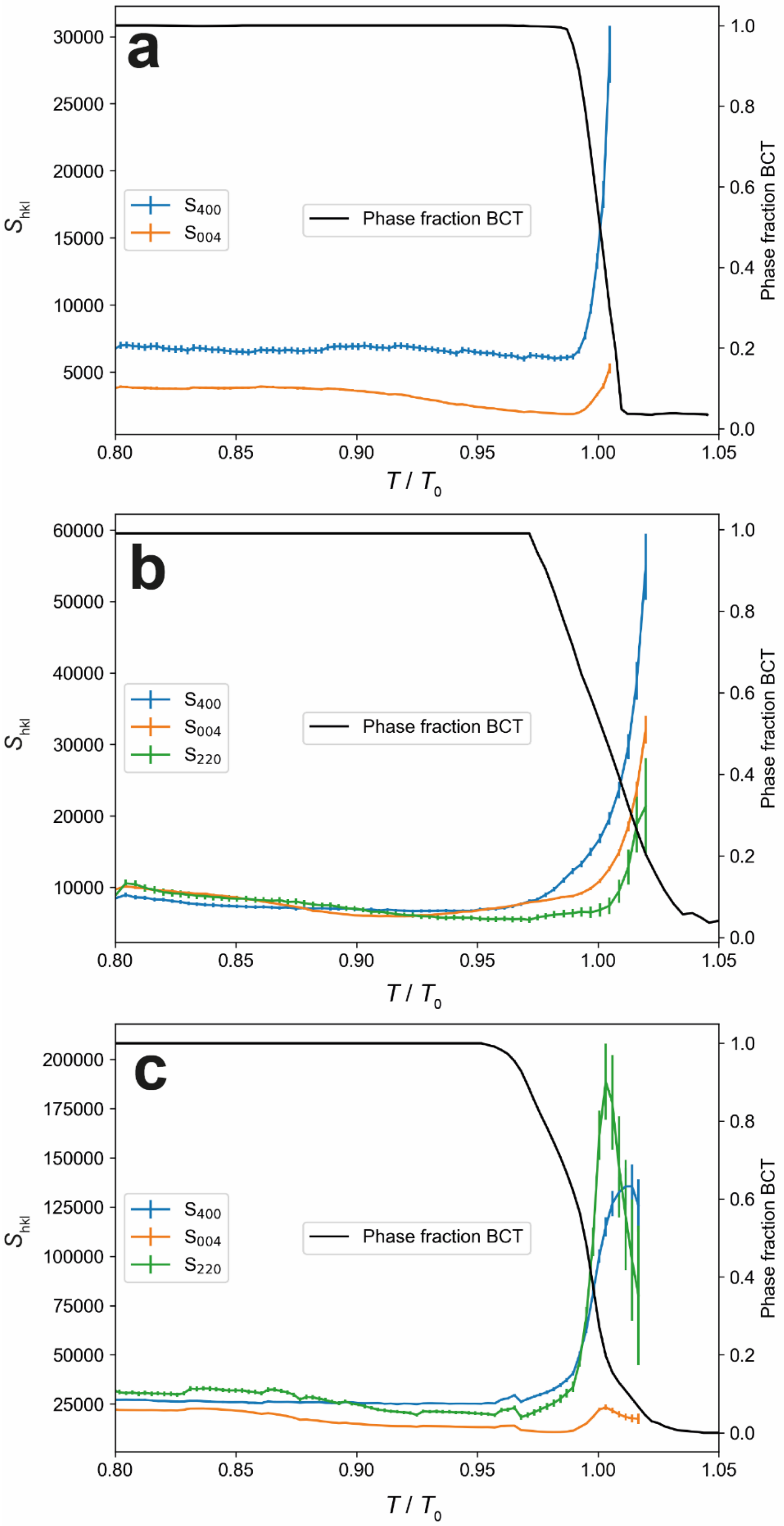

**Supplementary Fig. 15**: Anisotropic strain parameters ($S_{400}$, $S_{004}$ and $S_{220}$) obtained from the Rietveld refinements and BCT phase fraction for (a) PG and (b) NPG-PG and (c) NPG-PG-PE on heating. Refinements of PG including the $S_{220}$ parameter caused the fitting to become unstable, with negligible or negative values for $S_{220}$ being obtained. As such, it was not included in the final refinement for PG. NPG-PG-PE demonstrates proportionally increased strain associated with $S_{220}$ and $S_{400}$ as through the phase transition compared to the other two materials, indicating more disorder-induced broadening in the *ab* plane. We attribute this to disruption/weakening of the hydrogen bond network in the ternary solid solution. We attribute the variation in strain at low $T/T_0$ values as differences in crystal size and slight compositional variations in individual crystals as further evidenced in our electron diffraction data presented in Supplementary Information Note 1.

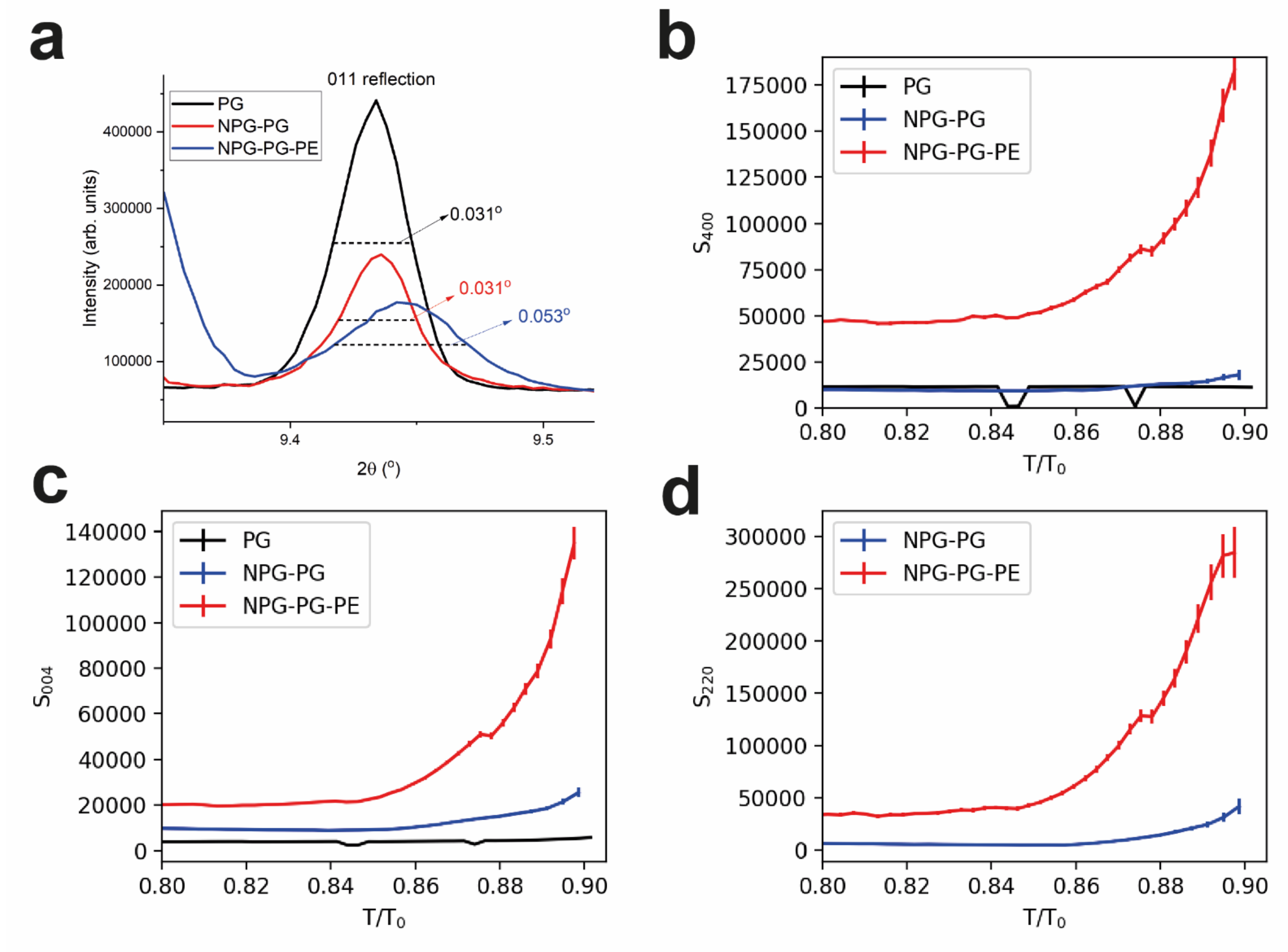

**Supplementary Fig. 16**: Analysis of peak broadening upon cooling below the PC-OC transition. (a) Representative full width half maximum (FWHM) values of the 011 reflection (in the BCT phase) for PG, NPG-PG and NPG-PG-PE on cooling. The datasets are collected at for $T/T_0$ = 1.00, where PG $T$ = 354K, NPG-PG $T$ = 302 K and NPG-PG-PE $T$ = 307 K. Despite the lower temperature, the NPG-PG-PE sample yields significantly broader peaks, which leads to the larger refined $S_{HKL}$ values. Other diffraction peaks show similar effects. (b) The $S_{400}$, (c) $S_{004}$ and (d) $S_{220}$ anisotropic strain parameters for cooling also show significantly larger changes for NPG-PG-PE close to the phase transition. Similar but much reduced trends are observed for NPG-PG and are interpreted as indicating greater disorder below $T_0$ in NPG-PG-PE, both on heating (Supplementary Fig. 15) and cooling. NB To refine the structure of NPG-PG and NPG-PG-PE from the diffraction data, it was necessary to include the $S_{220}$ broadening term at all temperatures. In contrast for PG, $S_{220}$ was found to be negligible and would not reliably refine to a stable value, demonstrating an enhanced broadening of *hk*0 type reflections in both binary and ternary compared to PG. We attribute the *h*00 and *hk*0 reflection broadening in the weakly first-order region of the transition, relative to the 00*l* reflection broadening, to destabilisation of the hydrogen bond network in the *ab* plane, facilitated by molecular substitution in the BCT crystal structure.

# Supplementary Information Note 1: Single crystal electron diffraction

All single crystal electron diffraction (3D-ED) measurements were obtained using a Rigaku XtaLAB Synergy-ED electron diffractometer, operated at 200 kV and equipped with a Rigaku HyPix-ED hybrid pixel array area detector. Powders (PG and NPG-PG-PE) were ground in a pestle and mortar and subsequently sprinkled on lacey carbon 200 mesh Cu grid. The grids were loaded into a Gatan Elsa cryo holder and cooled, to limit sample sublimation, from room temperature to 175 K. They were then transferred to the diffractometer and held under vacuum in the air lock chamber for a period for 30 minutes to ensure any ice had sublimated, then further cooled in in the column vacuum to 150 K. We investigated single crystal structures of both PG and NPG-PG-PE to compare with the compositional phase diagram presented in the main manuscript and our synchrotron powder x-ray diffraction results.

For PG, we find that the refined crystal structure agrees with our synchrotron powder XRD of adopting a BCT structure with an $I\bar{4}$ space group refined with a hydroxyl occupation set to 0.75 to match the expected composition. We obtain lattice parameters at 150 K of *a* = 6.0785 Å, *c* = 8.944 Å (see Table S1). These values are both ~1% larger than the values obtained from XRD at 283 K (see Supplementary Fig. 8). We anticipate the lattice constants obtained from ED to be less accurate than those from XRD, in part due to variations in the crystal height with respect to the eucentric height of the microscope. The full set of refined crystallographic data for PG can be obtained from Cambridge Crystallographic Data Centre (CCDC) under deposition number CCDC 2444733.

For NPG-PG-PE, we find that the refined crystal structure also agrees with our synchrotron powder XRD of adopting a BCT structure with an $I\bar{4}$ space group refined with a hydroxyl occupation set to 0.60 to match the expected composition. We obtain lattice parameters at 150 K of *a* = 6.0779 Å, *c* = 8.939 Å (see Table S2). Comparing this to the lattice parameters obtained from XRD at 246 K (see Supplementary Fig. 10) the *a* parameter is ~1.0% larger and the *c* parameter is ~0.3% smaller in ED. There may be some slight compositional variation between crystals of NPG-PG-PE which may account for some of the broadening of peaks observed in XRD for NPG-PG-PE when compared to PG, far from the phase transition (see Supplementary Fig. 15). The full set of refined crystallographic data for NPG-PG-PE can obtained from the CCDC under deposition number CCDC 2444734.

**Electron diffraction details for PG**

The crystal that was used for data collection is shown in Supplementary Fig. 17b. Data were acquired using rotation about a single axis between -60° to 50°, 0.25° image width with 440 frames at 3 s/°. Crystal data, data collection and structure refinement details are summarised in Table S1, Table S2 and Table S3, respectively. The asymmetric unit contains the central carbon atom (C1), C2 and 0.75 occupied O1 to account for three OH groups averaged over 4 crystallographically equivalent positions. Hydrogen atoms on C2 were modelled as 0.25 $CH_3$ and 0.75 $CH_2$ corresponding to the OH. The hydroxyl hydrogen atom was refined freely (position and isotropic ADP, Uiso), $CH_2$ hydrogens were placed in geometrically calculated positions and included as part of a riding model, the $CH_3$ hydrogens were included as part of a rigid rotor. C-H distances were set to standard distances from neutron diffraction and Uiso included as 1.2 and 1.5 times the parent carbon atom Ueq value for $CH_2$ and $CH_3$ groups respectively.

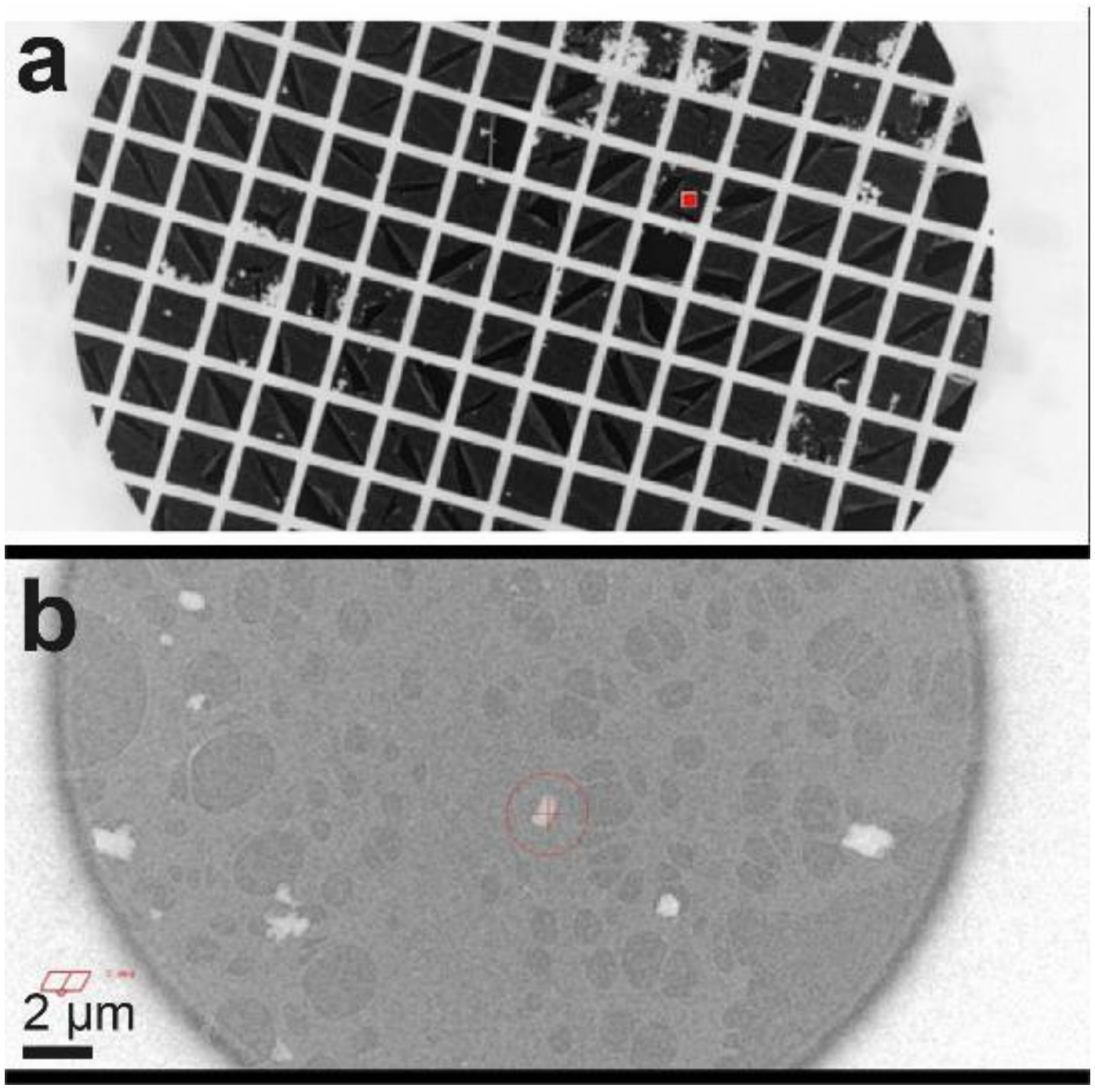

**Supplementary Fig. 17:** Electron microscopy images of (a) map of the lacy carbon grid with deposited PG powder (b) crystal used for data collection within the red circle, with a size of approximately 900 x 500 nm.

| | |
|---|---|
| $C_5H_{12}O_3$ | $D_x$ = 1.207 Mg m$^{-3}$ |
| $M_r$ = 120.15 | Electron radiation, l = 0.02510 Å |
| Tetragonal, I$\bar{4}$ | Cell parameters from 364 reflections |
| $a$ = 6.0795 (10) Å | q = 0.1–0.9° |
| $c$ = 8.944 (7) Å | m = 0.000 mm$^{-1}$ |
| $V$ = 330.6 (3)  Å$^3$ | $T$ = 150 K |
| $Z$ = 2 | Rectangular plate, white (bulk) |
| $F$(000) = 49.664 | |

**Table S1:** Refined PG crystal data.

| | |
|---|---|
| XtaLAB Synergy-ED, HyPix-ED, electron source at 200keV diffractometer | 340 independent reflections |
| Radiation source: electron diffractometer, JEOL JSM-2300ED, LaB6 | 205 reflections with $I$ ³ 2u($I$) |
| None monochromator | $R_{int}$ = 0.087 |
| Detector resolution: 10.0000 pixels mm$^{-1}$ | $q_{max}$ = 0.9°, $q_{min}$ = 0.1° |
| continuous rotation electron diffraction scans | $h$ = -7 to 7 |
| Absorption correction: multi-scan *CrysAlis PRO* 1.171.44.81a (Rigaku Oxford Diffraction, 2024) Empirical absorption correction using spherical harmonics, implemented in SCALE3 ABSPACK scaling algorithm. | $k$ = -7 to 7 |
| $T_{min}$ = 0.121, $T_{max}$ = 1.000 | $l$ = -11 to 11 |
| 781 measured reflections | |

**Table S2:** PG data collection information.

| | |
|---|---|
| Refinement on $F^2$ | 7 constraints |
| Least-squares matrix: full | Primary atom site location: dual |
| $R[F^2 > 2s(F^2)]$ = 0.106 | H atoms treated by a mixture of independent and constrained refinement |
| $wR(F^2)$ = 0.276 | $w = 1/[s^2(F_o^2) + (0.1215P)^2 + 0.0194P]$ where $P = (F_o^2 + 2F_c^2)/3$ |
| $S$ = 1.15 | $(D/s)_{max}$ = 0.001 |
| 340 reflections | $D\rho_{max}$ = 0.32 e Å$^{-3}$ |
| 27 parameters | $D\rho_{min}$ = -0.41 e Å$^{-3}$ |
| 9 restraints | |

**Table S3:** PG refinement information.

**Electron diffraction details for NPG-PG-PE**

The crystal that was used for data collection is shown in Supplementary Fig. 18b. Data were acquired using rotation about a single axis between -60° to -3.75°, 0.25° image width with 225 frames at 3 s/°. Crystal data, data collection and structure refinement details are summarised in Table S4, Table S5 and Table S6, respectively. The asymmetric unit contains the central carbon atom, C1, C2 and 0.60 occupied O1 to approximate the 60:38:02 composition of NPG-PG-PE as a 60:40 mixture of PG (three OH) and NPG (two OH) groups averaged over 4 crystallographically equivalent positions. Hydrogen atoms on C2 were modelled as 0.4 $CH_3$ and 0.6 $CH_2$ corresponding to the OH. The hydroxyl hydrogen atom was refined freely (position and isotropic ADP, Uiso), $CH_2$ hydrogens were placed in geometrically calculated positions and included as part of a riding model, the $CH_3$ hydrogens were included as part of a rigid rotor. C-H distances were set to standard distances from neutron diffraction and Uiso included as 1.2 and 1.5 times the parent carbon atom Ueq value for $CH_2$ and $CH_3$ groups, respectively.

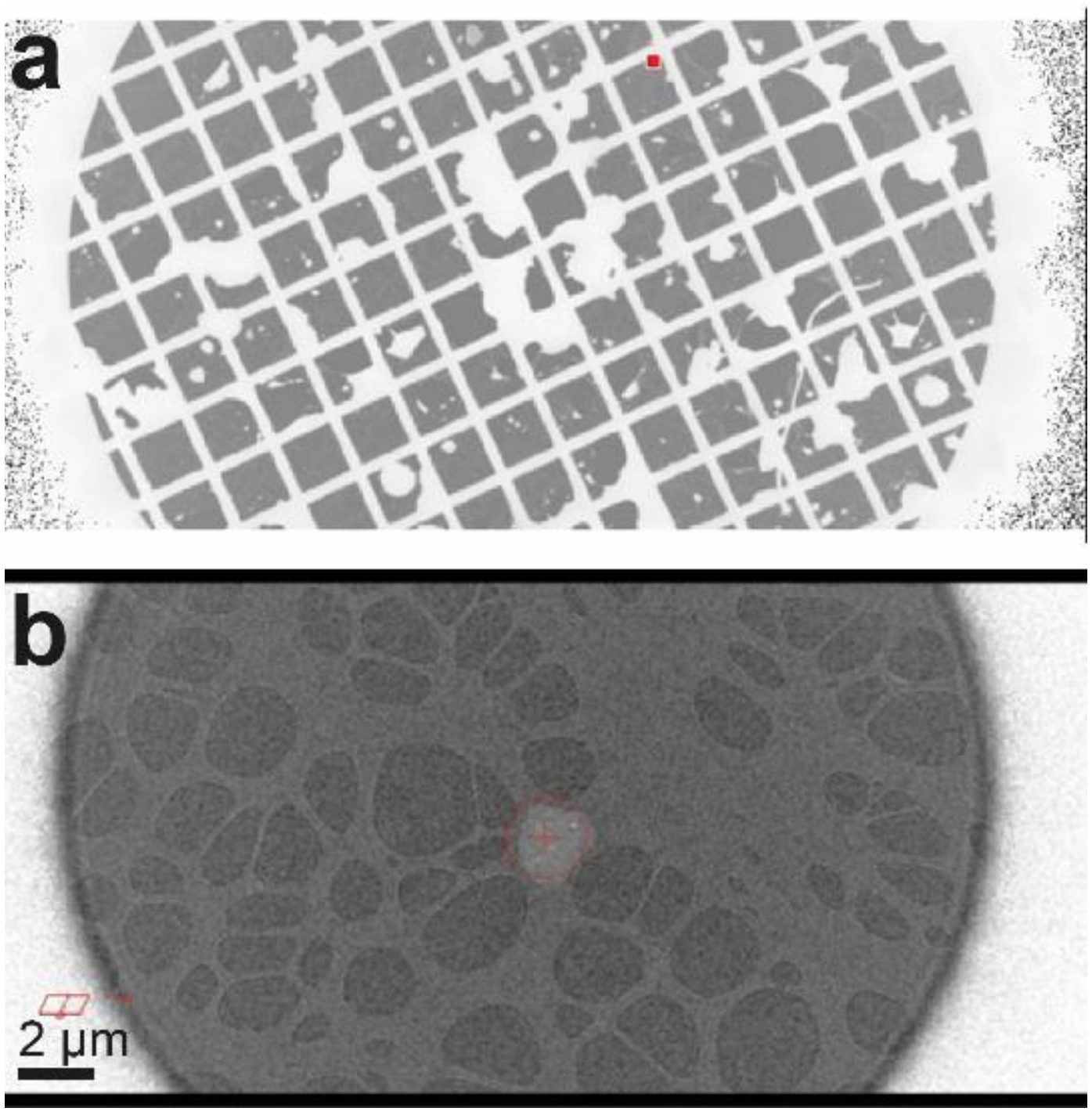

**Supplementary Fig. 18:** Electron microscopy images of (a) map of the lacy carbon grid with deposited NPG-PG-PE powder (b) crystal used for data collection within the red circle, with a size of approximately 2 x 2 μm.

| | |
|---|---|
| $C_5H_{12}O_{2.4}$ | $D_x$ = 1.112 Mg m$^{-3}$ |
| $M_r$ = 110.55 | Electron radiation, l = 0.02510 Å |
| Tetragonal, $I\bar{4}$ | Cell parameters from 256 reflections |
| *a* = 6.0779 (14) Å | q = 0.1–0.9° |
| *c* = 8.939 (3) Å | m = 0.000 mm$^{-1}$ |
| *V* = 330.20 (15) Å$^3$ | *T* = 150 K |
| *Z* = 2 | Plate, white (bulk) |
| *F*(000) = 47.284 | |

**Table S4:** Refined NPG-PG-PE crystal data.

| | |
|---|---|
| XtaLAB Synergy-ED, HyPix-ED, electron source at 200keV diffractometer | 252 independent reflections |
| Radiation source: electron diffractometer, JEOL JSM-2300ED, LaB6 | 211 reflections with $I > 2s(I)$ |
| None monochromator | $R_{int} = 0.025$ |
| Detector resolution: 10.0000 pixels $mm^{-1}$ | $q_{max} = 0.9°$, $q_{min} = 0.3°$ |
| continuous rotation electron diffraction scans | $h = -7$ to 7 |
| Absorption correction: multi-scan *CrysAlis PRO* 1.171.44.81a (Rigaku Oxford Diffraction, 2024) Empirical absorption correction using spherical harmonics, implemented in SCALE3 ABSPACK scaling algorithm. | $k = -4$ to 4 |
| $T_{min} = 0.140$, $T_{max} = 1.000$ | $l = -11$ to 11 |
| 398 measured reflections | |

**Table S5:** NPG-PG-PE data collection information.

| | |
|---|---|
| Refinement on $F^2$ | 9 constraints |
| Least-squares matrix: full | Primary atom site location: dual |
| $R[F^2 > 2s(F^2)] = 0.182$ | H-atom parameters constrained |
| $wR(F^2) = 0.490$ | $w = 1/[s^2(F_o^2) + (0.2P)^2]$ where $P = (F_o^2 + 2F_c^2)/3$ |
| $S = 2.19$ | $(D/s)_{max} = -0.001$ |
| 252 reflections | $D\rho_{max} = 0.20$ e Å$^{-3}$ |
| 24 parameters | $D\rho_{min} = -0.15$ e Å$^{-3}$ |
| 9 restraints | |

**Table S6:** NPG-PG-PE refinement information.

### Electron diffraction computing details

Data collection: *CrysAlis PRO* system (CCD 44.81a 64-bit (release 22-11-2024)); cell refinement: *CrysAlis PRO* 1.171.44.81a; data reduction: *CrysAlis PRO* 1.171.44.81a; program(s) used to solve structure: SHELXT 2018/2 [2]; program(s) used to refine structure: olex2.refine 1.5 [3]; software used to prepare material for publication: Olex2 1.5 [4].

# Supplementary Information Note 2: XRD phase fraction fitting

In order to fit only the IFWS data that corresponds to temperatures where the material is fully in the PC phase we first parameterise the FCC phase fraction on heating ($\Phi_{\mathrm{FCC,heat}}$) and cooling ($\Phi_{\mathrm{FCC,cool}}$). This is achieved by fitting error functions to the full heating and cooling synchrotron XRD data resulting in:

$$\Phi_{\mathrm{FCC,heat}} = \frac{1}{2}\mathrm{erf}\big(0.124(T - T_0)\big) + \frac{1}{2} \quad \text{(S1)}$$

and

$$\Phi_{\mathrm{FCC,cool}} = \frac{1}{2}\mathrm{erf}\big(0.151(T - T_{\mathrm{SC}})\big) + \frac{1}{2} \quad \text{(S2)}$$

where the numerical argument in the error functions indicates the slope of the error function, showing that, as expected, the cooling transition occurs over a narrower temperature range than the heating transition. The above equations can then be used to determine the temperatures at which $\Phi_{\mathrm{FCC}}$ approaches unity on both heating and cooling.

# Supplementary Information Note 3: QENS Data Analysis & Fitting

The measured QENS data, $S_{\text{measured}}(Q,E)$ comes from a convolution of the instrumental resolution function, $R(Q,E)$, and the theoretical scattering function, $S(Q,E)$, through[5]:

$$S_{\text{measured}}(Q,E) = R(Q,E) \otimes [S(Q,E)] + B(Q), \tag{S3}$$

where

$$S(Q,E) = A_0(Q)\delta(E) + \sum_{i=1}^{n} A_i L_i. \tag{S4}$$

These equations were used directly as the model function for QENS, representing an elastic peak ($A_0(Q)\delta(E)$), one ($n$ = 1) or two ($n$ = 2) Lorentzian components ($A_i L_i$) and a flat background ($B(Q)$), all convoluted with the measured resolution function of the instrument ($R(Q,E)$).

Simultaneous global fitting of FWS and QENS data was performed to separate the behaviour of the modes above the phase transition as we have implemented previously for pure NPG[6]. Four datasets were included in the fit: IFWS on cooling with ±3 μeV and ±6 μeV energy transfer windows, and QENS scans at 330 K and 350 K, each containing spectra for 16 different $Q$ bins.

For the IFWS data, the above equation was modified by removing the elastic contribution and modelling a temperature dependence by using an Arrhenius law for the linewidths, $\Gamma_i(Q)$:

$$\Gamma_i(Q) = \Gamma_{0_i}(Q) e^{-\frac{E_{a_i}}{k_B T}} \tag{S5}$$

Subsequently, parameter ties were established between the different datasets to achieve the following:

- a global, $Q$-independent value for the activation energy of each process ($E_{a_1}$ and $E_{a_2}$);
- consistency between the IFWS linewidths, calculated using the Arrhenius law above, and the fitted Lorentzian linewidths $\Gamma_i(Q)$ from the QENS spectra at different temperatures; and
- identical ratios of the amplitudes of Lorentzian contributions ($A_1(Q)$ / $A_2(Q)$) across all spectra (QENS and IFWS) with the same $Q$.

To calculate the elastic incoherent structure factor (EISF) of each mode above the phase transition we separated the slow and fast components. The EISF of the slow component was approximated according to:

$$\text{EISF}_{\text{slow}}(Q) = \frac{I_{\text{elastic}}(Q)}{I_{\text{elastic}}(Q) + I_{\text{inelastic,slow}}(Q)} \tag{S6}$$

Likewise, we determined the EISF of the fast component according to:

$$\text{EISF}_{\text{fast}}(Q) = \frac{I_{\text{elastic}}(Q) + I_{\text{inelastic,slow}}(Q)}{I_{\text{elastic}}(Q) + I_{\text{inelastic,slow}}(Q) + I_{\text{inelastic,fast}}(Q)} \tag{S7}$$

To account for the possibility that only a fraction of scatterers participates in the observed process with a given geometry, we modelled a fractional $\text{EISF}_{\text{observed}}(Q)$:

$$\text{EISF}_{\text{observed}}(Q) = (1-f) + f\,\text{EISF}_{\text{model}}(Q) \tag{S8}$$

where $f$ is the fraction of scatterers involved in the motion [7–9].

The methyl rotation (Mode 1) is well described by the 120° three hop (3-Hop) model, which takes the form[5]:

$$\mathrm{EISF}_{3-\mathrm{Hop}}(Q) = (1-f) + f\,\frac{1}{3}\left[1 + \frac{2\sin(\sqrt{3}Qr)}{\sqrt{3}Qr}\right] \qquad \text{(S9)}$$

where $r$ is the rotational radius of $^1$H motion associated with this mode. We model hydroxymethyl rotation (Mode 2) as the $^1$H scatterers hopping between equivalent sites around a circle. The EISF for this continuous hop (C-Hop) is taken as:

$$\mathrm{EISF}_{\mathrm{C-Hop}}(Q) = (1-f) + f\,\frac{1}{6}\left[1 + 2\frac{\sin(Qr)}{Qr} + 2\frac{\sin(\sqrt{3}Qr)}{\sqrt{3}Qr} + \frac{\sin(2Qr)}{2Qr}\right]. \qquad \text{(S10)}$$

For ease, equation S10 describes the case for hopping between 6 equivalent sites on a circle of radius $r$, as it has been shown that this is a good approximation for more than 6 hopping sites within the $Q$-range observed here. Finally, the EISF associated with molecular reorientation (Mode 3) describes $^1$H atoms continuously diffusing on a sphere with radius $r$:

$$\mathrm{EISF}_{\mathrm{reorientation}}(Q) = (1-f) + f\left[\frac{\sin(Qr)}{Qr}\right]^2. \qquad \text{(S11)}$$

Based on the composition of NPG-PG-PE (60%, 38%, 2%), which defines the number of $^1$H atoms in the methyl and hydroxymethyl groups, and assuming the above models well describe the rotational motions, we would expect Mode 1 to have an $f$ of 0.605, Mode 2 to have an $f$ of 0.395 and Mode 3 to have an $f$ of 1.0.

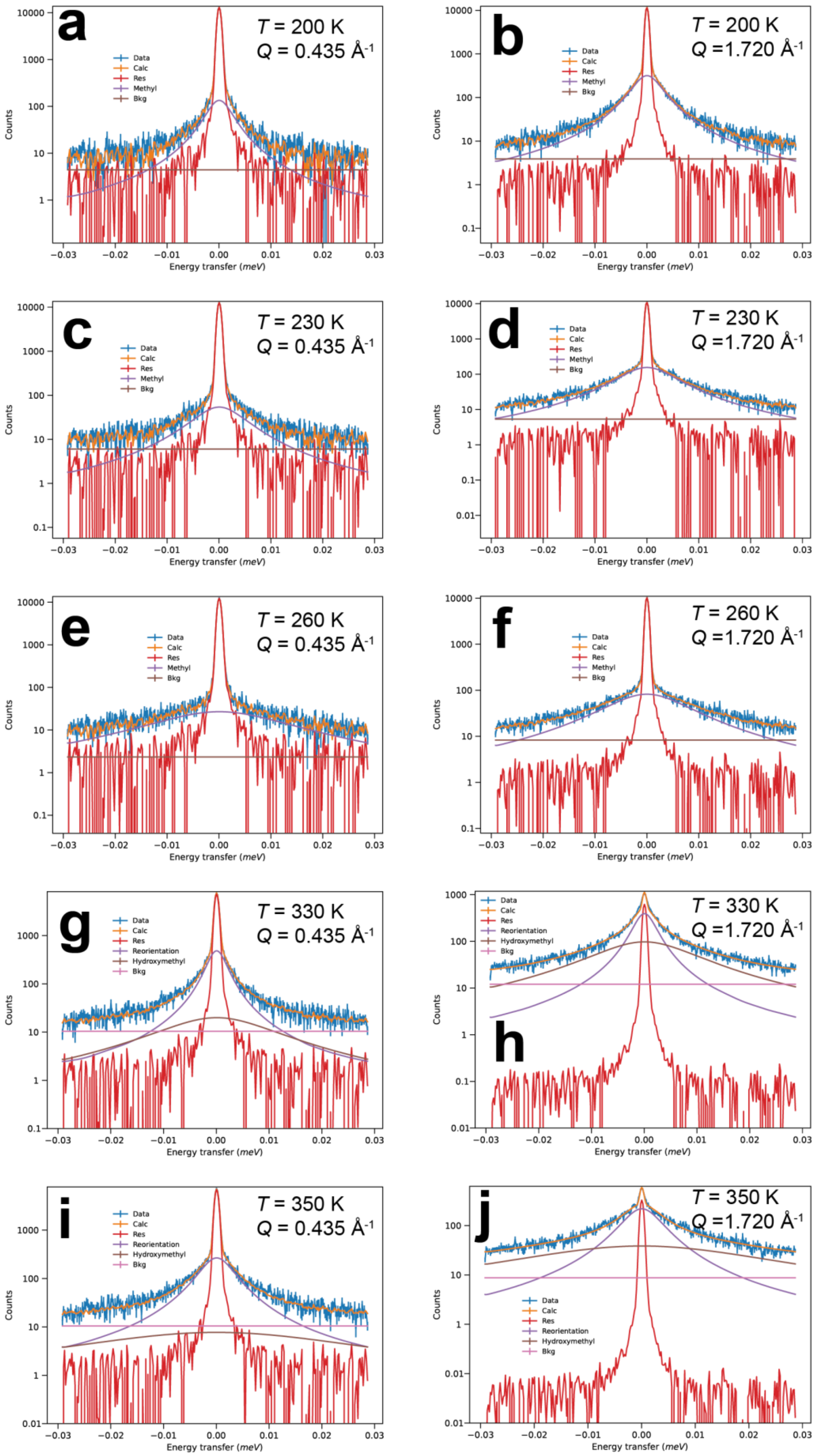

**Supplementary Fig. 19**: QENS global fits at (a-i) low (0.435 Å$^{-1}$) and (b-j) high (1.720 Å$^{-1}$) *Q* values at all temperatures measured. The intensity ratio of the two modes (reorientation and hydroxymethyl) at ±3 μeV constrained the amplitude ratio of the IFWS Lorentzians presented in Supplementary Fig. 21, during global fitting as defined in Supplementary Information Note 3.

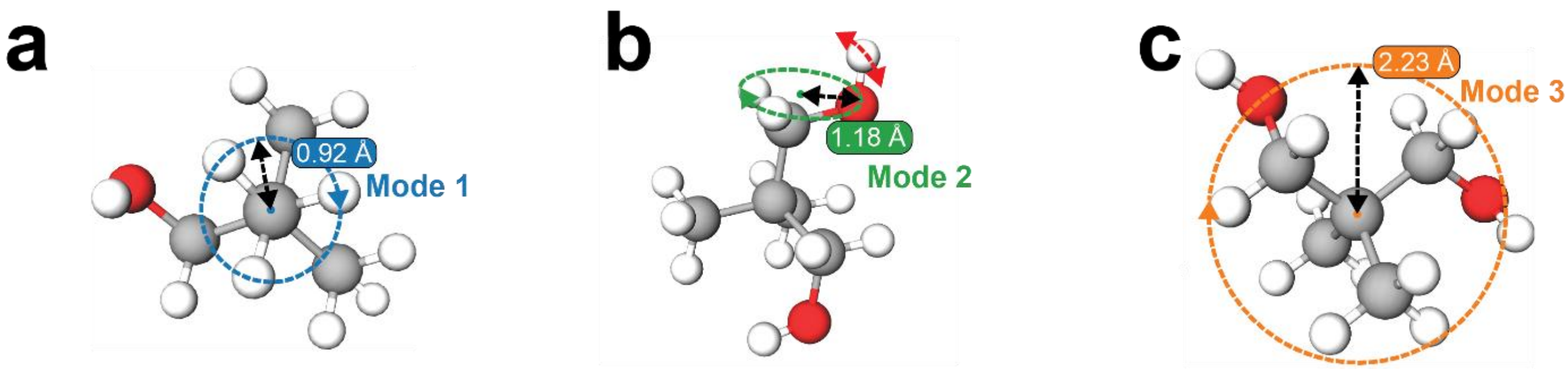

**Supplementary Fig. 20**: Schematics of the NPG molecule, indicating the geometric origin for the radii of rotation obtained from the EISF fits in the main text. (a) methyl rotation, (b) hydroxymethyl rotation and (c) molecular reorientation. Figure has been reproduced from our previously published work[6].

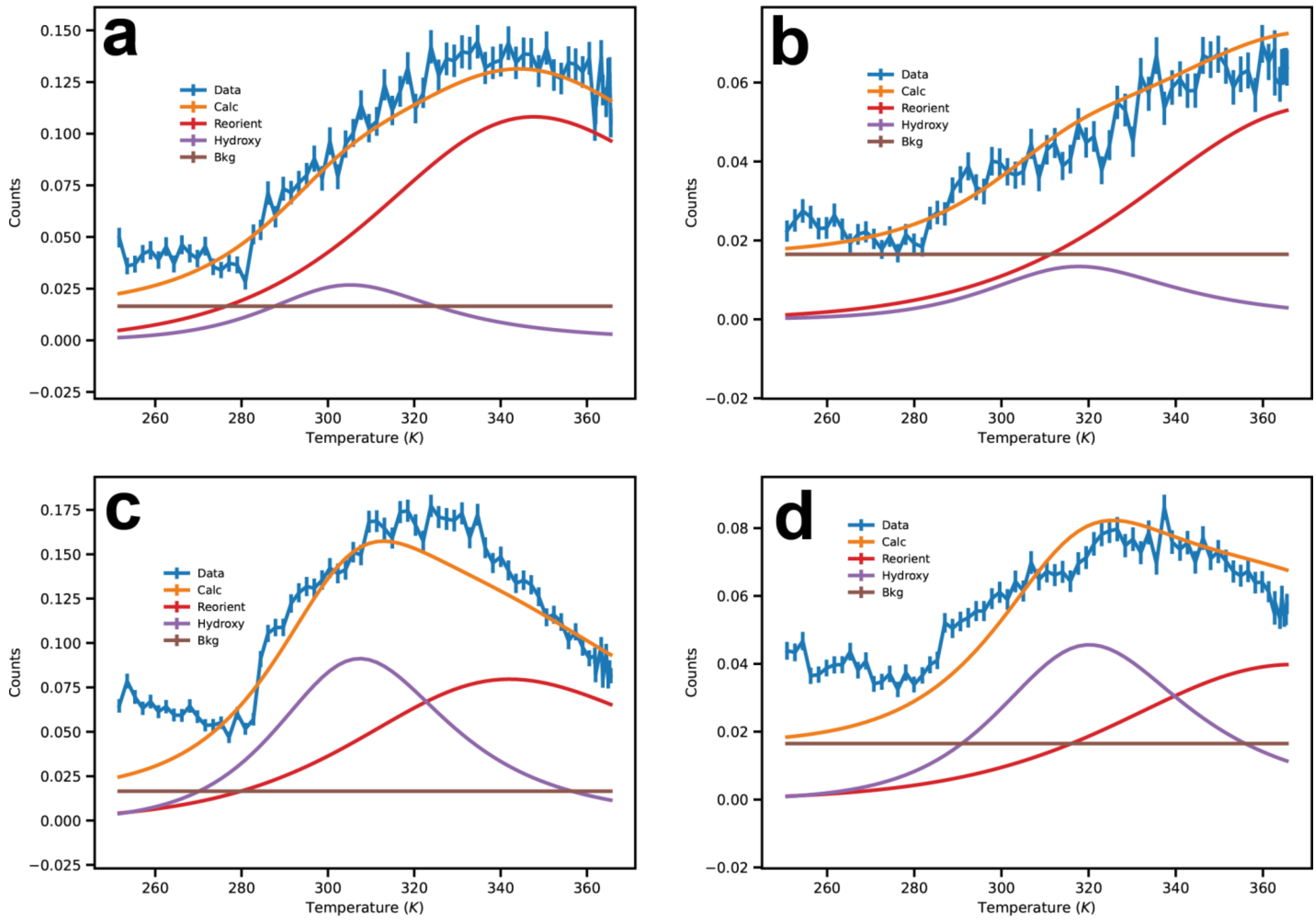

**Supplementary Fig. 21**: IFWS on cooling global fits for (a) $Q$ = 0.44 Å$^{-1}$, $E$ = 3 μeV (b) $Q$ = 0.44 Å$^{-1}$, $E$ = 6 μeV (c) $Q$ = 1.72 Å$^{-1}$, $E$ = 3 μeV (d) $Q$ = 1.72 Å$^{-1}$, $E$ = 6 μeV. Each plot shows the two high temperature modes: hydroxymethyl rotation (hydroxy) and molecular reorientation (reorient). Constraints with QENS data presented in Supplementary Fig. 19, during global fitting as defined in Supplementary Information Note 3. Data below 293 K was excluded from the fits since that region corresponds to the phase transition and the low temperature OC phase. The flat background was fixed to the average IFWS intensity for $T$ < 100 K.

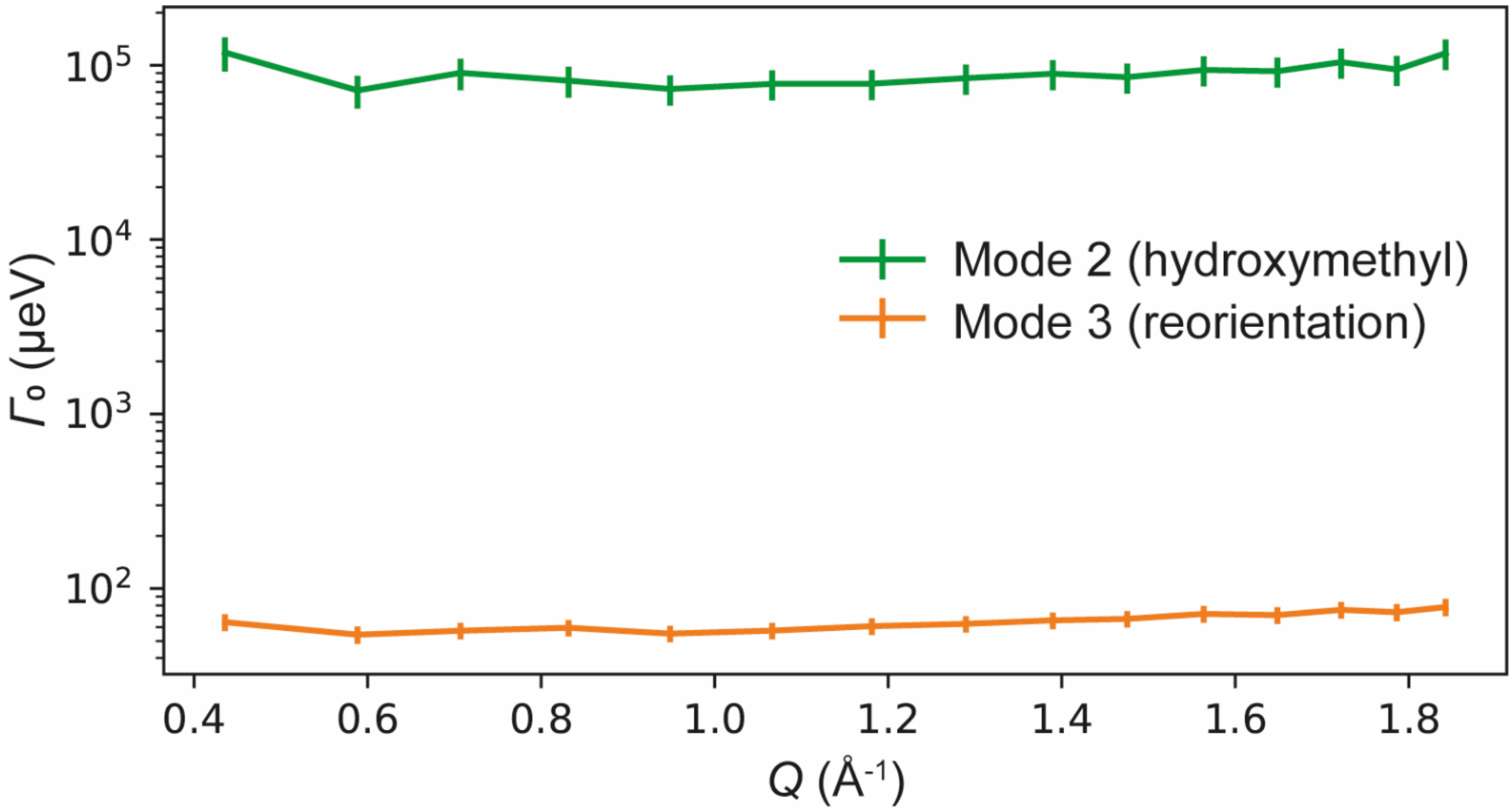

**Supplementary Fig. 22**: Pre-exponential factors ($\Gamma_0(Q)$), which gives the half width half maximum (HWHM) linewidth of the respective mode at infinite temperature, obtained from IFWS fitting of cooling data. The mean values (across all *Q*) are 90 ± 4 meV for hydroxymethyl rotation (Mode 2, green) and 65 ± 2 µeV for molecular reorientation (Mode 3, orange). Converting the pre-exponential factors to attempt frequencies using the relation $\tau = \frac{\hbar}{\Gamma}$ provides an estimate for the attempt frequencies of the dynamic processes: 140 THz (Mode 2) and 100 GHz (Mode 3).

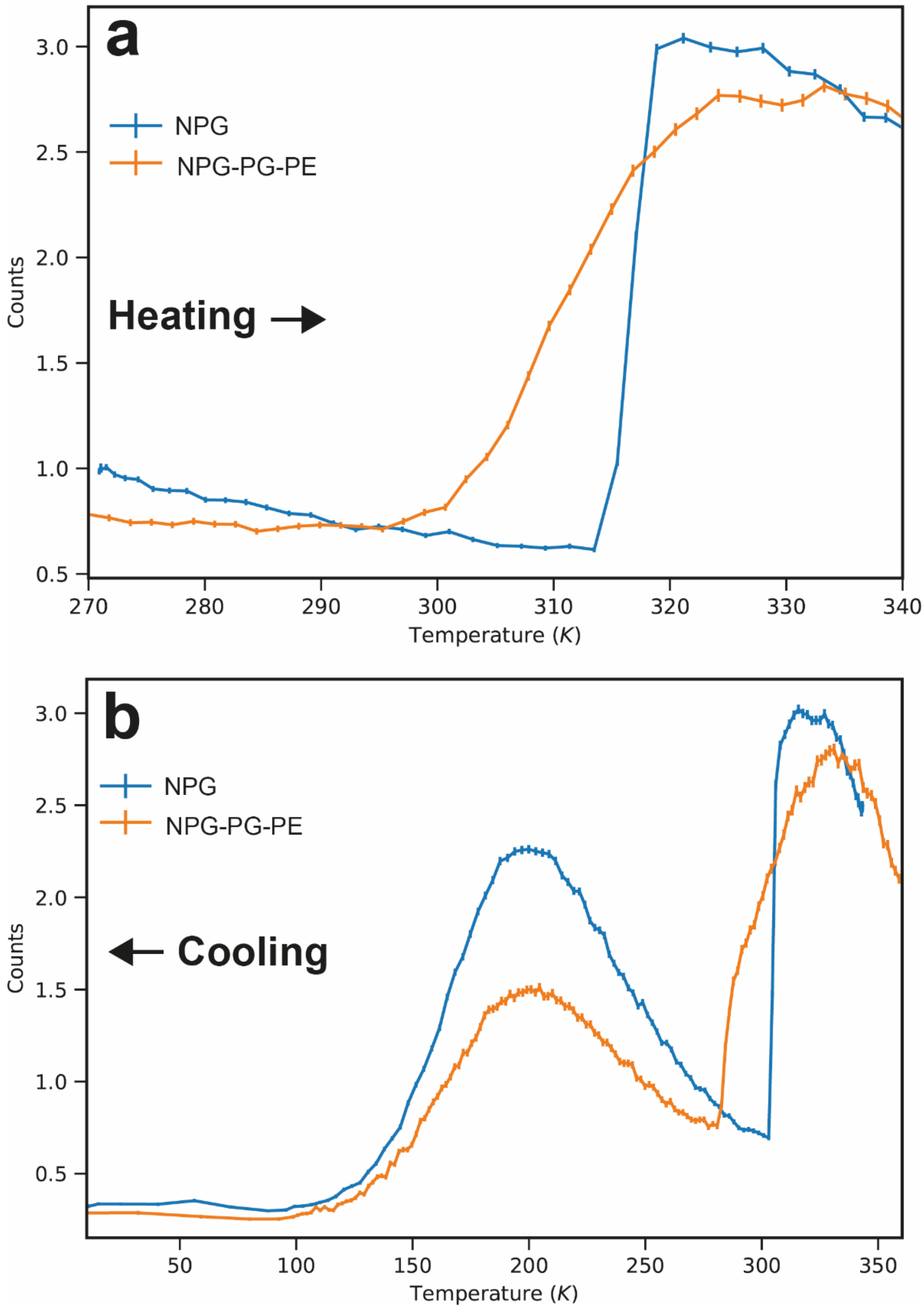

**Supplementary Fig. 23**: IFWS with ±3 μeV energy windows, summed across all measured *Q* (0.19 to 1.89 $Å^{-1}$) compared between NPG and NPG-PG-PE on (a) heating and (b) cooling. Changes in intensity here correspond to dynamic modes "moving through" the energy window. On heating(cooling) intensity increases(decreases) as broadening of the elastic line increases(decreases). The large phase coexistence region of NPG-PG-PE compared to NPG is clearly seen in (a). In (b), the intensity difference between the peaks centred at 200 K, which corresponds to methyl rotation, is due to the different number of methyl groups present in NPG-PG-PE compared to NPG.

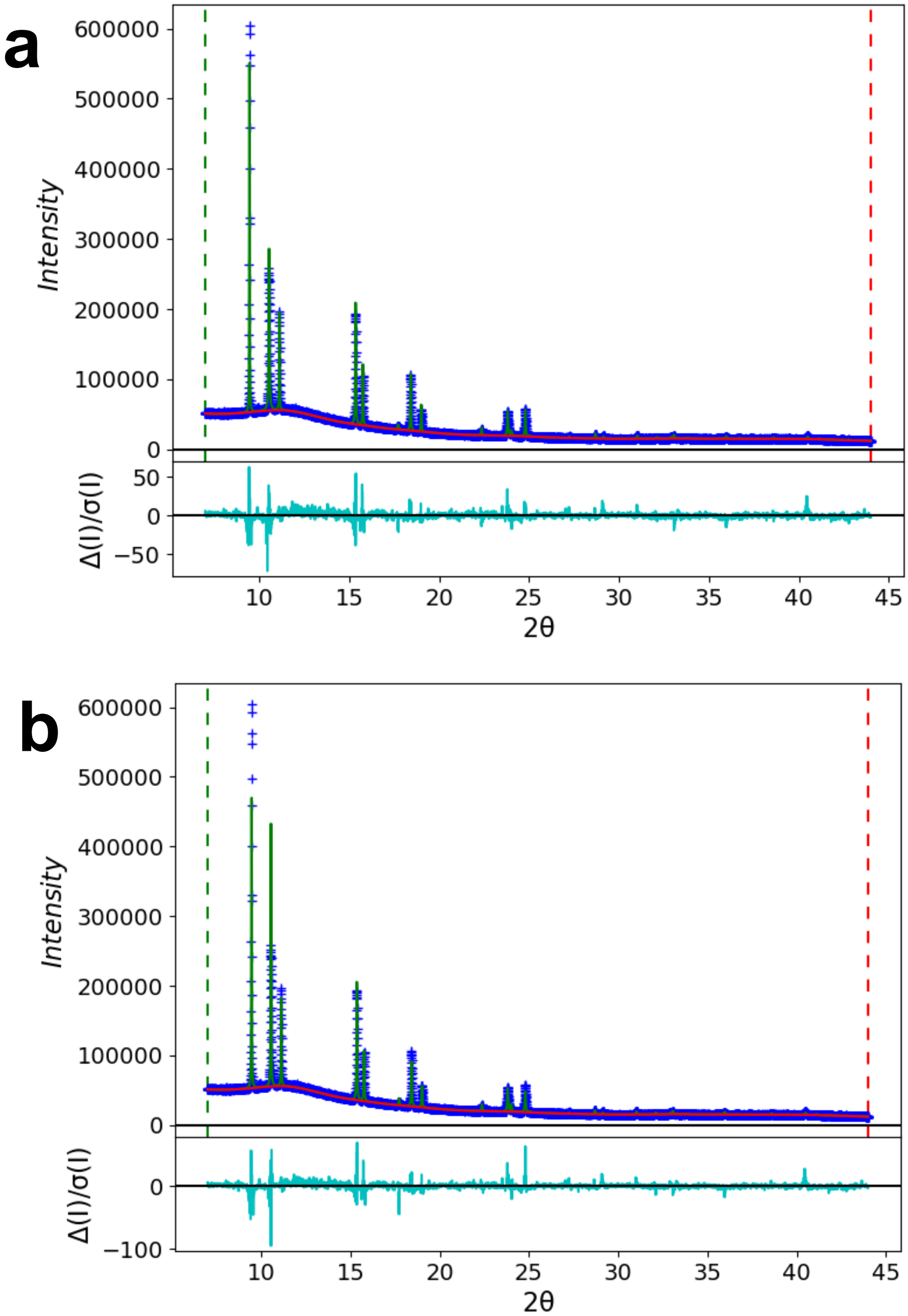

**Supplementary Fig. 24**: A comparison between the use of (a) anisotropic and (b) isotropic broadening models for the refinement of NPG-PG, demonstrating that the anisotropic strain model significantly improves the refinement. (a) Rietveld refinement of NPG-PG at 240K with Stephen's anisotropic broadening. *wR* = 2.821; Reduced $\chi^2$ = 11.07; $N_{obs}$ = 9251; $N_{vals}$ = 8. (b) Same as data as in (a) but with isotropic microstrain broadening (not Stephen's anisotropic broadening model). *wR* = 3.225; Reduced $\chi^2$= 14.45. $N_{obs}$ = 9251; $N_{vals}$ = 5. Note the scale of $\Delta(I)/\sigma(I)$ in both graphs and the significantly improved fits to the 011 and 002 reflections at 9.465° and 10.553˚, respectively.